\documentclass{aa}  
\usepackage{widetext}
\usepackage[draft]{pgf}
\usepackage{graphicx}
\usepackage{txfonts}
\usepackage{color}
\usepackage{gensymb}
\usepackage{soul}
\usepackage{txfonts}
\usepackage{color}
\usepackage{soul}
\usepackage{booktabs}
\usepackage{multirow}
\usepackage{mathtools}
\usepackage{lipsum}
\def\teff{T_{\text{eff}}}
\def\stage{t_{\star}}
\def\xv{\boldsymbol{x}}
\def\yv{\boldsymbol{y}}
\def\muv{\boldsymbol{\mu}}
\def\epsilonv{\boldsymbol{\epsilon}}
\def\thetav{\boldsymbol{\theta}}
\def\thetasv{\boldsymbol{\theta_S}}
\def\betasv{\boldsymbol{\beta_S}}
\def\thetastv{\boldsymbol{\theta_{\star}}}

\def\Bv{\boldsymbol{\mathcal{B}}}
\def\Kz{\mathcal{K}_0}
\def\Ko{\mathcal{K}_1}
\def\Xstv{\boldsymbol{X_{\star}}}
\def\Sigmav{\boldsymbol{\Sigma}}

\begin{document}

   \title{Latitudinal differential rotation in the solar \\ analogues 16 Cyg A and B}


   \author{
     M. Bazot\inst{1,2}
     \and
     O. Benomar\inst{2}
     \and
     J. Christensen-Dalsgaard\inst{3}
     \and
     L. Gizon\inst{2,4,5}
     \and
     S. Hanasoge\inst{2,6}
     \and
     M. Nielsen\inst{2}
     \and
     P. Petit\inst{7}
     \and
     K.~R.~Sreenivasan\inst{2}
   }

   \institute{Division of Sciences, New York University Abu Dhabi, United Arab Emirates\email{mb6215@nyu.edu}
     \and
     Center for Space Science, NYUAD Institute, New York University Abu Dhabi, PO Box 129188, Abu Dhabi, United Arab Emirates
     \and
     Stellar Astrophysics Centre, Department of Physics and Astronomy, Aarhus University, Ny Munkegade 120, DK-8000 Aarhus C, Denmark
     \and
     Max-Planck-Institut f\"ur Sonnensystemforschung, G\"ottingen, Germany
     \and
     Institut f\"ur Astrophysik, Georg-August-Universit\"at G\"ottingen, G\"ottingen, Germany
     \and
     Department of Astronomy \& Astrophysics, Tata Institute of Fundamental Research, Mumbai 400005, India
     \and
     Institut de Recherche en Astrophysique et Plan\'etologie, Universit\'e de Toulouse UPS--OMP / CNRS, 14, avenue \'Edouard Belin, 31400 Toulouse, France
             }

   \date{Received September , ; accepted March , }

 
  \abstract
  {Asteroseismology has undergone a profound transformation as a scientific field following the CoRoT and \emph{Kepler} space missions. The latter is now yielding the first measurements of latitudinal differential rotation obtained directly from oscillation frequencies. Differential rotation is a fundamental mechanism of the stellar dynamo effect.}
   {Our goal is to measure the amount of differential rotation in the solar analogues 16 Cyg A and B, which are the components of a binary system. These stars are the brightest observed by \emph{Kepler} and have therefore been extensively observed, with exquisite precision on their oscillation frequencies.}
   {We modelled the acoustic power spectrum of 16 Cyg A and B using a model that takes into account the contribution of differential rotation to the rotational frequency splitting. The estimation was carried out in a Bayesian setting. We then inverted these results to obtain the rotation profile of both stars under the assumption of a solar-like functional form.}
   {We observe that the magnitude of latitudinal differential rotation has a strong chance of being solar-like for both stars, their rotation rates being higher at the equator than at the pole. The measured latitudinal differential rotation, defined as the difference of rotation rate between the equator and the pole, is $320\pm269$~nHz and $440^{+363}_{-383}$~nHz for 16 Cyg A and B, respectively, confirming that the  rotation rates of these stars are almost solar-like. Their equatorial rotation rates are $535\pm75$~nHz and $565_{-129}^{+150}$~nHz. Our results are in good agreement with measurements obtained from spectropolarimetry, spectroscopy, and photometry.}
   {We present the first conclusive measurement of latitudinal differential rotation for solar analogues. Their rotational profiles are very close to those of the Sun. These results depend weakly on the uncertainties of the stellar parameters.}


   \maketitle
%

\section{Introduction}

Rotation is a ubiquitous characteristic of stars; it has many connections to convection, stellar pulsations, and magnetic fields \citep[e.g.][]{Tassoul00}. It can also be related to the evolution of close-by planets through tidal effects \citep[e.g.][]{Privitera16}. Despite this central role that it plays in stellar physics, a full understanding of its behaviour remains elusive. Only for the Sun are measurements precise enough to provide a proper insight on rotation \citep[e.g.][]{Thompson03}, owing to the large number of observed solar pulsation eigenmodes \citep{Hill96}. The spherical-harmonic degrees of the detected modes reach values as high as $l\simeq200$ at low radial orders \citep{Larson08}. Higher values, up to $\gtrsim 1500$, can be reached by fitting ridges in the power spectrum \citep{Couvidat13}. In a spherically symmetric, non-rotating star, the eigenfrequencies of the non-radial modes are degenerate with respect to the azimuthal order. However, perturbations to the velocity field caused by the rotational flow can lift this degeneracy. The resulting frequency splitting is related to the rotational flow by a linear integral equation. It is possible to invert this relation to obtain information on the solar rotation rate \citep[see e.g. ][]{Thompson91}. In contrast, the degrees of the eigenmodes observed in other stars are typically in the range $l = 0-2$, with some observations detecting $l=3$ modes \citep{Bouchy01,Bazot12,Appourchaux12,Metcalfe12}. As a consequence, inverting the eigenfrequencies becomes far more challenging. 

There nevertheless exist sources of information on rotation in stars other than Sun. Spectroscopic measurements offer estimates of the surface velocity as it broadens absorption lines through the Doppler effect. In general, the velocity is known up to a $\sin i$ factor, with $i$ the inclination of the spin axis of the star with respect to the observer's line of sight; this cannot be estimated from spectroscopic measurements alone. Recently, however, progress has been made through the advent of asteroseismology, and in particular, through the space missions CoRoT \citep{Baglin09} and \emph{Kepler} \citep{Borucki10}.

Improvements have come in two ways. First, it is now possible to measure frequency splittings with precision \citep{Gizon13,Appourchaux14,Nielsen14}. These relate to the average rotation rate of a star. Furthermore, the inferred value of the rotational splitting is geometrically tied to the inclination of the star \citep{Gizon03} through its effect on the observed mode amplitudes. Adequate estimates of the former therefore require estimates of the latter. Combined with spectroscopic measurements and radius determinations, splitting and inclination determinations have provided consistency checks of the asteroseismically estimated surface rotational velocity \citep{Gizon13}. It has also been argued that discrepancies between the two measurements can be interpreted as a signature of radial differential rotation \citep{Benomar15,Deheuvels15}.

The other major source of information on stellar rotation rates comes from photometric light curves and is related to stellar activity. It is well known that spots and plages transiting the stellar surface induce drops and increases in the integrated flux, respectively. If the activity signal is sufficiently long and coherent, this produces a quasi-periodic modulation of the intensity that is correlated with the stellar rotation rate. Such signatures have been extensively studied for large samples of \emph{Kepler} stars \citep{Reinhold13b,Walkowicz13,Garcia14,McQuillan14,Nielsen13}. Some caution is needed with the interpretation of these rotation rates because they depend on the latitude of the spots through the variation with latitude of the surface rotation rate.

Latitudinal differential rotation is indeed a major characteristic of the solar rotation profile. It is well established that the Sun rotates faster at the equator than at the poles by a factor of roughly 1.4. It has also been shown \citep{Schou98,Chaplin99,Thompson03} that this latitudinal differential rotation remains approximately constant as a function of radius throughout the convective zone. In the radiative zone, at least outside its inner 20\%, the rotation rate becomes that of a solid body. For this reason, latitudinal differential rotation is believed to be the result of the interplay between rotation and convective flows.

Much uncertainty currently remains as to which convective scale is the main driver of this phenomenon. On the one hand, it has been speculated that a mean-field approach of turbulent convection can explain differential rotation. The basic picture consists of describing the convective flow as a stationary component plus a time-dependent turbulent term that after insertion in the Navier-Stokes equations for a rotating fluid, gives rise to Reynolds stresses. The latter is in turn related to the radial and latitudinal gradients of the rotation rates, that is, to differential rotation \citep{Ruediger74,Ruediger80,Ruediger82}. This type of model can reproduce the observations made for the Sun in a rather satisfactory way \citep{Kitchatinov95,Kitchatinov17}. On the other hand, the global approach consists of solving the full set of the equations of hydrodynamics using the proper spherical coordinates and boundary conditions. This work was initiated by \citet{Gilman81}. Given the spherical arrangement of the system, differential rotation is caused by the Coriolis force acting on large-scale convective motion \citep{Tassoul00}.  Recent work has allowed modelling enough spatial convective scales so that these simulations can reproducethe solar observations to some extent \citep{Guerrero13,Guerrero16}. These two explanations need not be mutually exclusive. Finally, hydrodynamic simulations for different stellar masses and rotation rates have been carried out \citep[see e.g.][]{Brun17}. These show a wide range of morphologies for rotation flows in Sun-like stars. It is therefore clear that observing differential rotation is key to understanding the interplay between convection and rotation. It is our goal to provide such information for Sun-like stars.

Our refined picture of differential rotation in the Sun is due to the large number of modes observed in this star, which greatly helps the inversion process. The turning point of a given mode in the stellar interior depends on its degree. Consequently, the wide range of $l$ values obtained for the Sun allows us to probe many layers of its internal structure. 
Until recently, the relatively low number of frequencies detected in other stars (typically about 20 modes) had been an obstacle to inversion. Information on latitudinal differential rotation was gathered from three main sources. First, it is possible to invert the relation described above between the spot rotation rates and differential rotation to infer the latter \citep{Donahue96,Reinhold13a,Lanza14,Davenport15}. This is done through spot modelling \citep[e.g.][]{Dumusque14}, which involves an implicit physical model for the stellar spots. In the same spirit, some studies have tried to identify variations in the measurements of CaII emission lines, which are tied to active stellar regions \citep{Bertello12}. Interesting peculiar cases of photometric spot modelling are encountered for planet-hosting stars when the companion transits in front of stellar spots. The resulting decrease in transit depth may allow a precise characterisation of differential rotation \citep{Valio17}. Another method involves the use of Doppler maps obtained from spectroscopy or spectropolarimetry \citep[e.g.][]{Donati97a,Marsden11}. The last approach available consists of analysing the Fourier transforms of spectral absorption lines, whose side lobes are sensitive to the variation in latitude of the rotation rate \citep{Huang61,Gray77}. This strategy has been used extensively to detect differential rotation in A- and F-type stars \citep{Reiners02a,Reiners02b,Reiners03,AvE12}. In a recent study, however, \citet{Benomar18} demonstrated the possibility of measuring the magnitude of differential rotation in 13 Sun-like stars using the techniques of asteroseismology on \emph{Kepler} time series. We extend this work here.

We report seismic detection of latitudinal differential rotation for 16 Cyg A and B. These stars have physical characteristics close to those of the Sun; in particular, the rotation rates obtained from spectroscopy suggest roughly solar values \citep{Takeda05}. For this reason, they have sometimes been dubbed {\lq}solar analogues{\rq}. It is therefore of prime interest to assess whether the measured latitudinal differential rotation is also close to solar-like or if a difference exists. Most of the stars with measured differential rotation reported in \citet{Benomar18} are leaning towards the F type. A handful of them have effective temperatures closer to the Sun, but still hotter by roughly 300~K. Therefore, 16 Cyg A and B are the best candidates to study rotation under nearly solar conditions, which is supported by the occasional classification of these stars as {\lq}solar twins{\rq} (e.g. \citealp{King97} ; see also the introduction of \citealp{Bazot18} for a note on the class of solar twins).

 Another important fact about 16~Cyg~A and B is that they are some of the brightest stars that the \emph{Kepler} mission observed. Therefore, their oscillation frequencies were estimated with extremely good precision \citep{Metcalfe12,Davies15}. This gives us an opportunity to address the problem of how the estimates of latitudinal differential rotation depend on the uncertainties in the stellar parameters (mass, age, initial chemical composition, and mixing-length parameter).

In Section~\ref{sect:spectrum} we describe the general procedure for data fitting, including the modelling of differential rotation effects. We present its outcome in terms of coefficient estimates for a functional expansion of the rotational splitting. In Section~\ref{sect:inversion} we explain the inversion methods used for both stars, with an emphasis on the more difficult case of 16~Cyg~B. We discuss in Section~\ref{sect:blabla} the impact  of the uncertainties on the stellar parameters on our results. We also explore the implication of the measured stellar deformation on the magnetic field of the stars.

\section{Acoustic mode fitting with differential rotation}\label{sect:spectrum}
\begin{figure*}
   \centering
   \includegraphics[width=\textwidth]{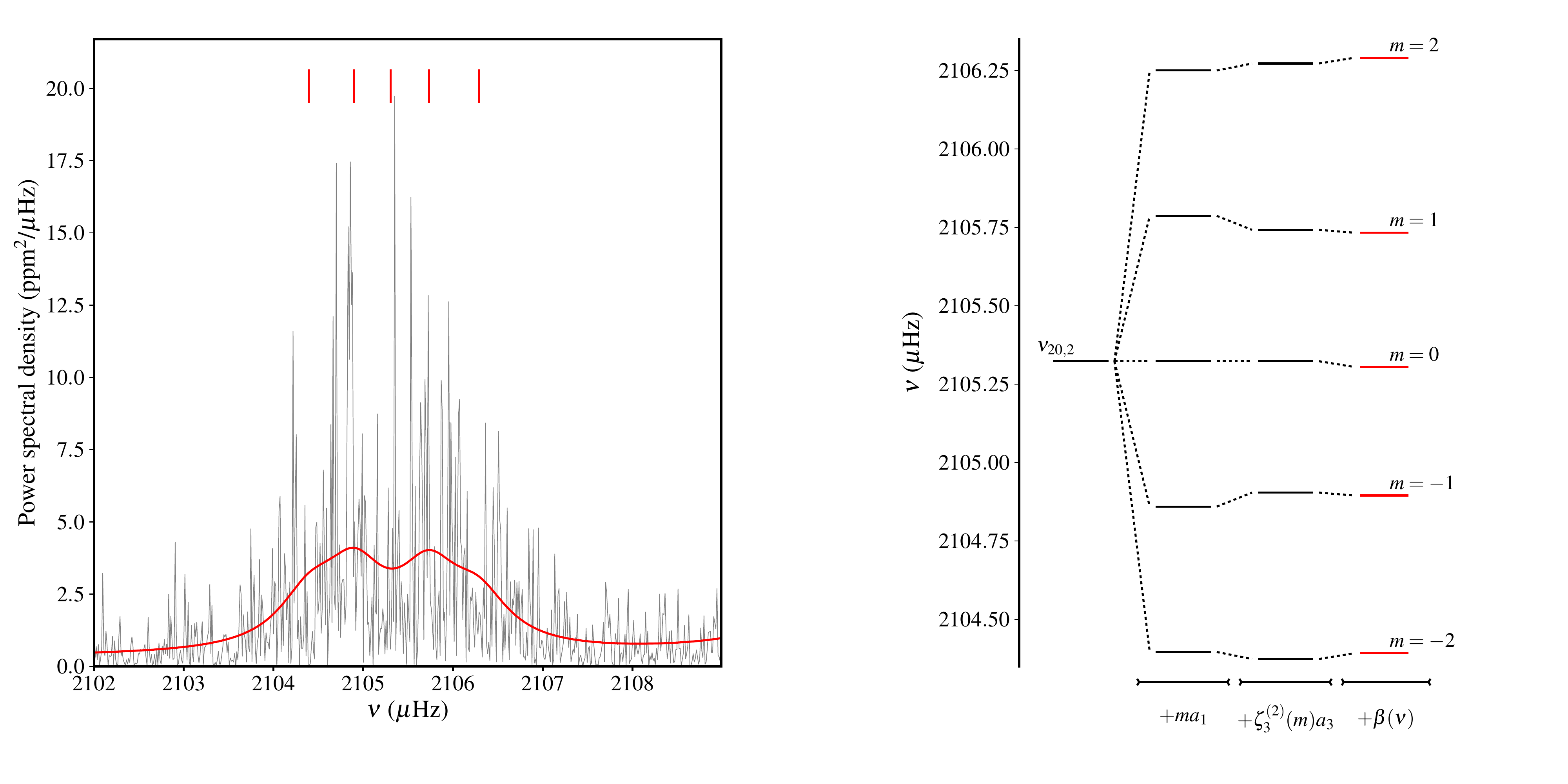}
   \caption{{\bf Left panel}: Spectrum of 16~Cyg~A in the region of the $(l,n) = (2,20)$ multiplet. Black shows the observations and red is the theoretical model corresponding to the MAP estimates of the parameters $\thetav$ (cf. Eq. \ref{eq:param}). The vertical red ticks mark the position of the non-degenerate frequencies with $-l\leq m \leq l$. {\bf Right panel}: Splitting diagram for the multiplet. The contributions of the terms in Eq.~(\ref{eq:fdist}) are represented individually. The final red horizontal ticks correspond to those in the left panel.}
         \label{fig:spectrum-16CygA}
\end{figure*}

\subsection{Data}

The sources 16 Cyg A and B have been observed by \emph{Kepler} from 13 September 2010 to 8 March 2013 (covering \emph{Kepler} quarters Q7 to Q16). Their magnitudes,$V=5.95$ and 6.20, respectively place them beyond the saturation limit of the on-board CCDs. For this reason, it was necessary to produce a specific pixel mask that allows measuring the flux with fewer pixels \citep{Metcalfe12}. The raw data were treated using the procedure described in \citet{Garcia11}. It corrects for the instrumental perturbations (outliers, jumps, drifts, etc.) and merges time series for observations spanning multiple quarters.

The resulting precision on the measured flux, and subsequently, on the derived oscillation frequencies, are the best obtained by \emph{Kepler,} and to this day, for any Sun-like star in addition to the Sun itself. Overall, 54 and 56 modes have been detected for16 Cyg A and B \citep{Davies15}, respectively, with the precision ranging from 0.03~$\mu$Hz to 2.74~$\mu$Hz. Interestingly, \citet{Davies15} reported projected rotational-splitting measurements of $411\pm 13$~nHz for 16 Cyg A and $274\pm 17$~nHz for 16 Cyg B , that is, measurements that include a $\sin i$ factor,. However, they did not detect conclusive signatures of differential rotation. Our goal is to remodel these acoustic power spectra with a model that differs from those used in \citet{Metcalfe12} and \citet{Davies15} in that it takes differential rotation into account. We use the same time series as in \citet{Davies15}.

\subsection{Spectrum model}

The fitting of the power spectra obtained from these times series is based on principles that are commonly adopted in asteroseismic studies. The first assumption is that the power spectrum in each frequency bin we consider in the Fourier space is independent of (and therefore not correlated to) its neighbours; that is, we neglect leakage coming from the convolution of the Fourier transforms of the signal and the window function. This allows us to consider the probability density of the power in each frequency bin separately. Making the further assumption that the noise on the measurements is Gaussian, the power spectrum $P_i = P(\omega_i)$ at any frequency $\omega_i$ is exponentially distributed, with a probability density
\begin{equation}\label{eq:expdist}
\displaystyle
f(P_k) = \frac{1}{\mathcal{P}(\omega_k)} \exp\left(-\frac{P_k}{\mathcal{P}(\omega_k)}\right),\end{equation}
with $\mathcal{P}(\omega_k)$ the average value of the power spectrum at $\omega_k$, where $k$ is an index of the frequency bins. A common model for the average power spectrum is a sum of Lorentzian functions centred at the eigenfrequencies of the pulsation modes. This is suitable for regularly damped and re-excited modes such as those observed in the Sun \citep{Anderson90}. The central frequencies, the widths, and the heights of the Lorentzian are free parameters of the spectrum model. Usually, multiplets resulting from a rotationally induced lifting of degeneracy are fit jointly using a relation of the form $\nu_{n,l,m} = f(\nu_{n,l} , m ; (a_j)_{1\leq j\leq J})$. Here $\nu_{n,l}$ is the {\textquotedblleft}central{\textquotedblright} frequency of the multiplet, of radial order $n$ and angular degree $l$, $m$ is the azimuthal order, and $(a_j)_{1\leq j\leq J}$ is a vector of coefficients allowing us to expand the splitting \citep{Ritzwoller91} as
\begin{equation}\label{eq:splitting}
  \displaystyle
  \delta\nu_{n,l,m} = \nu_{n,l,m} - \nu_{n,l} = \sum_{j=1}^J a_j(n,l)\zeta^{(l)}_j(m),
\end{equation}
with the functions $\zeta^{(l)}_j(m)$ forming an orthogonal basis such that $\sum\limits_m \zeta^{(l)}_j(m)\zeta^{(l)}_i(m) = 0$ for $i\neq j$. 

In the expression of $\mathcal{P}(\omega_k)$ we introduce the effect of differential rotation. It is typical in asteroseismology to retain only the first-order term in the above expansion. The $a_1$ coefficient can be interpreted as a weighted average of the rotation throughout the star \citep{Appourchaux14,Davies15}. In this work, however, we also consider the next term in Eq.~(\ref{eq:splitting}), as suggested by \citet{Gizon04}. This leads to a frequency distribution described by
\begin{equation}\label{eq:fdist}
  \displaystyle
  \nu_{n,l,m} = \nu_{n,l} + ma_1(n,l) + \beta_{n,l,m}(\nu) + \zeta_3^{(l)}(m)a_3(n,l).
\end{equation}
In the following we only consider splittings for $l=2$, therefore we use $\zeta_3^{(2)}(m) = (5m^3 - 17m)/3$. We also note that a frequency-dependent term, $\beta_{n,l,m}(\nu)$, has been added to the expression resulting from Eq.~(\ref{eq:splitting}). It includes the perturbation to the frequencies stemming from the asphericities of the star. They may be caused by the centrifugal forces, perhaps also by a large-scale magnetic field \citep{Gough90}, a tidal distortion, or a strong anisotropic stellar wind. We show in Fig.~\ref{fig:spectrum-16CygA} a typical multiplet, chosen at $n=20$, $l = 2$, as observed in the power spectrum of 16~Cyg~A and modelled using Eq.~(\ref{eq:fdist}). The effects of the higher-order term and departure from sphericity are smaller than the contribution of $a_1$ to the splitting. This is shown in the right panel of Fig.~\ref{fig:spectrum-16CygA}, where we represent the individual contributions to the non-degenerate frequencies. It is interesting to note that $ma_1$ and $\zeta_3^{(l)}(m)a_3$ are symmetric functions of the azimuthal order $m$, while $\beta_{n,l,m}$ is antisymmetric in $m$. 

The coefficients $a_1$ and $a_3$ now become parameters of the model to be fitted to the observed spectrum. The relative heights of the modes in a multiplet also depend on the inclination $i$ \citep{Gizon03}. We denote the parameters necessary to describe the spectrum as 
\begin{equation}
\thetav = (\thetasv,a_1,a_3,i,\betasv,\Bv) \; .
\label{eq:param}
\end{equation}
Here $\thetasv\in \mathbb{R}^{3N}$ is a vector grouping of $\nu_{n,l}$, $H_{n,l}$ , and $\Gamma_{n,l}$, which are the frequency, height, and width of the Lorentzian describing the expectation value of the line profile of mode $(n,l)$ in the power spectrum. $N$ oscillation frequencies have been observed. Other parameters are $a_1$, $a_3$, the stellar inclination $i \in [0,2\pi]$, and $\betasv$, a vector grouping of the parameters we used in the functional form of $\beta_{n,l,m}$. We also used a model to describe the noise contribution to the power spectrum. Its parameters are collectively denoted as $\Bv,$ and we refer to \citet{Benomar09} for a discussion of the issues of background-noise fitting in seismic spectra.

These parameters were estimated in a Bayesian framework, that is, we estimated the posterior density function of the parameter vector $\thetav$, conditional on $\yv$, the data
\begin{equation}\label{eq:posterior}
  \displaystyle
  \pi_{\thetav\vert\yv}(\thetav\vert\yv) \propto \pi_{\thetav}(\thetav)L(\thetav).
\end{equation}
Here $\yv = (P_1,\dots,P_K)$ is the observation vector. We always note the probability density of a variable $x$ by $\pi_x$. Likewise, if the probability is conditional on $\yv$, for instance, the corresponding density is noted $\pi_{x\vert\yv}$.

The likelihood function $L$ is the product of the individual exponential distributions for all the bins considered. 
\begin{equation}\label{eq:likelihood}
  \displaystyle
  L(\theta) = \prod_{k=1}^{K}\frac{1}{\mathcal{P}(\omega_k ; \theta)} \exp\left(-\frac{P_k}{\mathcal{P}(\omega_k ; \theta)}\right).
\end{equation}

We recall that this function is the probability density of the data when $\yv$ is the variable and is called the likelihood when seen as a function of the parameters, $\thetav$. In that case, it is not a probability density of the parameters.

The prior $\pi_{\theta}$ adopted for the parameters can be decomposed as the product of priors on the individual parameters by assuming that they are independent of each other. The priors on the usual parameters (frequencies, mode heights, and line widths) are described in the Appendix of \citet{White16}. In addition, we chose uniform priors on $a_1$ and $a_3$. The coefficient $a_1$ was assumed to be positive. Furthermore, because we do not expect the stars to be fast rotators, an upper boundary was set to 5~$\mu$Hz. The boundaries for the prior on $a_3$ were more difficult to set, and we used a test-and-trial procedure. We finally set $-0.1$~$\mu$Hz~$\leq a_3 \leq $~0.1~$\mu$Hz. We chose the following functional form for $\beta_{n,l,m}$ (see also Sect.~\ref{sect:blabla}):
\begin{equation}\label{eq:beta}
\beta_{n,l,m}(\nu_{n,l,m}) = \beta_0Q_{l,m}\nu_{n,l}.
\end{equation}
Here we have $Q_{l,m} = [L^2 - 3m^2]/[(2l -1)(2l + 3)]$, with $L^2 = l(l+1)$ \citep{Gough90,Kjeldsen98}, and $\betasv$ reduces to the scalar $\beta_0$.

The corresponding posterior density was sampled using an adaptive Markov chain Monte Carlo algorithm based on the scheme described by \citet{Haario01}, with some modifications inspired by \citet{Atchade06}. The simulations were run for $1\,000\,000$ iterations. We used a burn-in sequence of $100\,000$ samples. In Fig.~\ref{fig:spectrum-16CygA} we represent part of the spectrum obtained for the values of the parameters corresponding to the global maximum of $\pi_{\theta\vert\yv}$, that is, the maximum a posteriori (MAP).

\begin{figure}
   \centering
   \includegraphics[width=.55\textwidth]{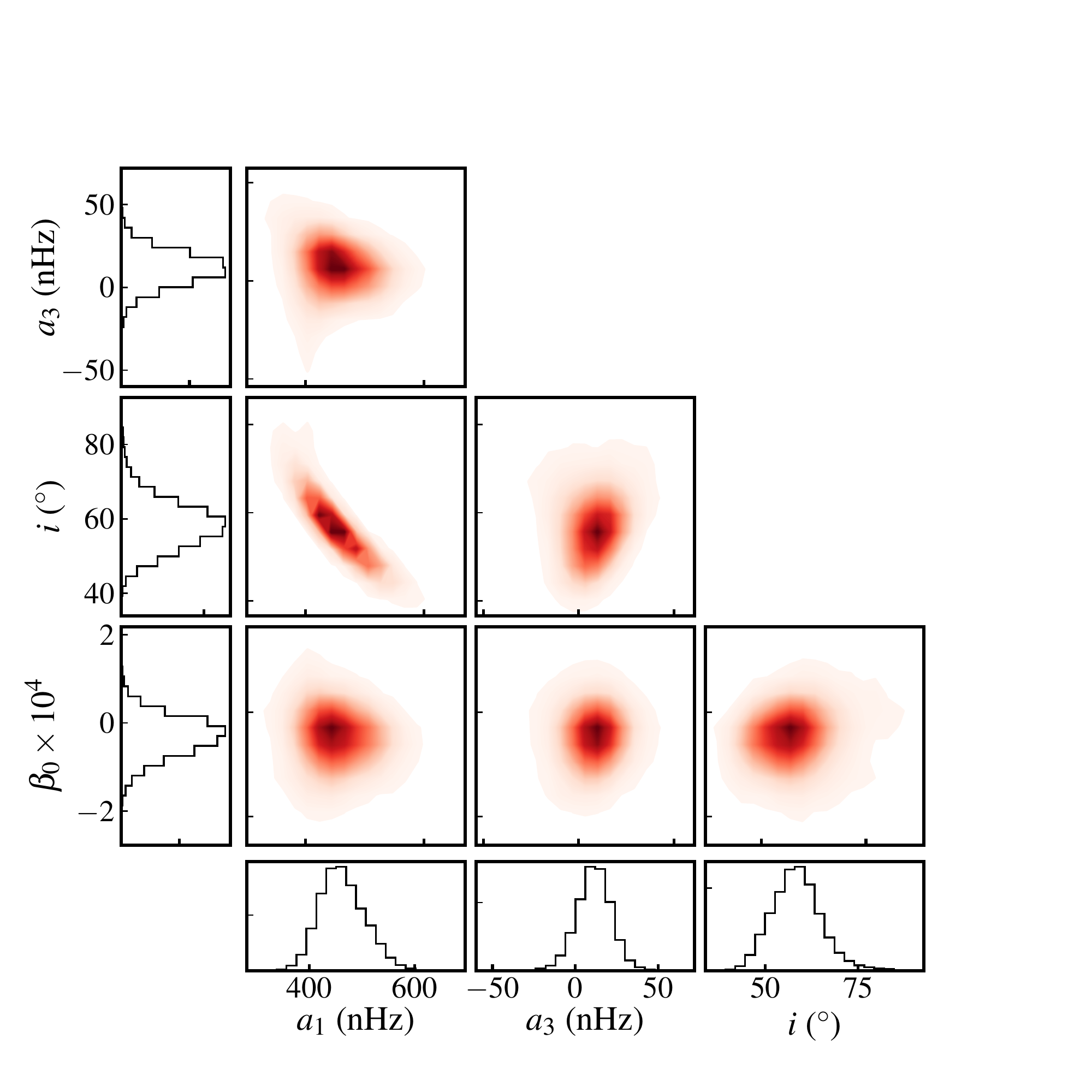}
   \caption{Marginal densities for the spectrum parameters $a_1$, $a_3$, $i,$ and $\beta_0$ of 16~Cyg~A. The central panels show the joint marginal densities of the paired parameters. In the side panels we plot the individual marginal densities. 
   }
         \label{fig:joint-16CygA}
   \end{figure}

\begin{figure}
   \centering
   \includegraphics[width=.55\textwidth]{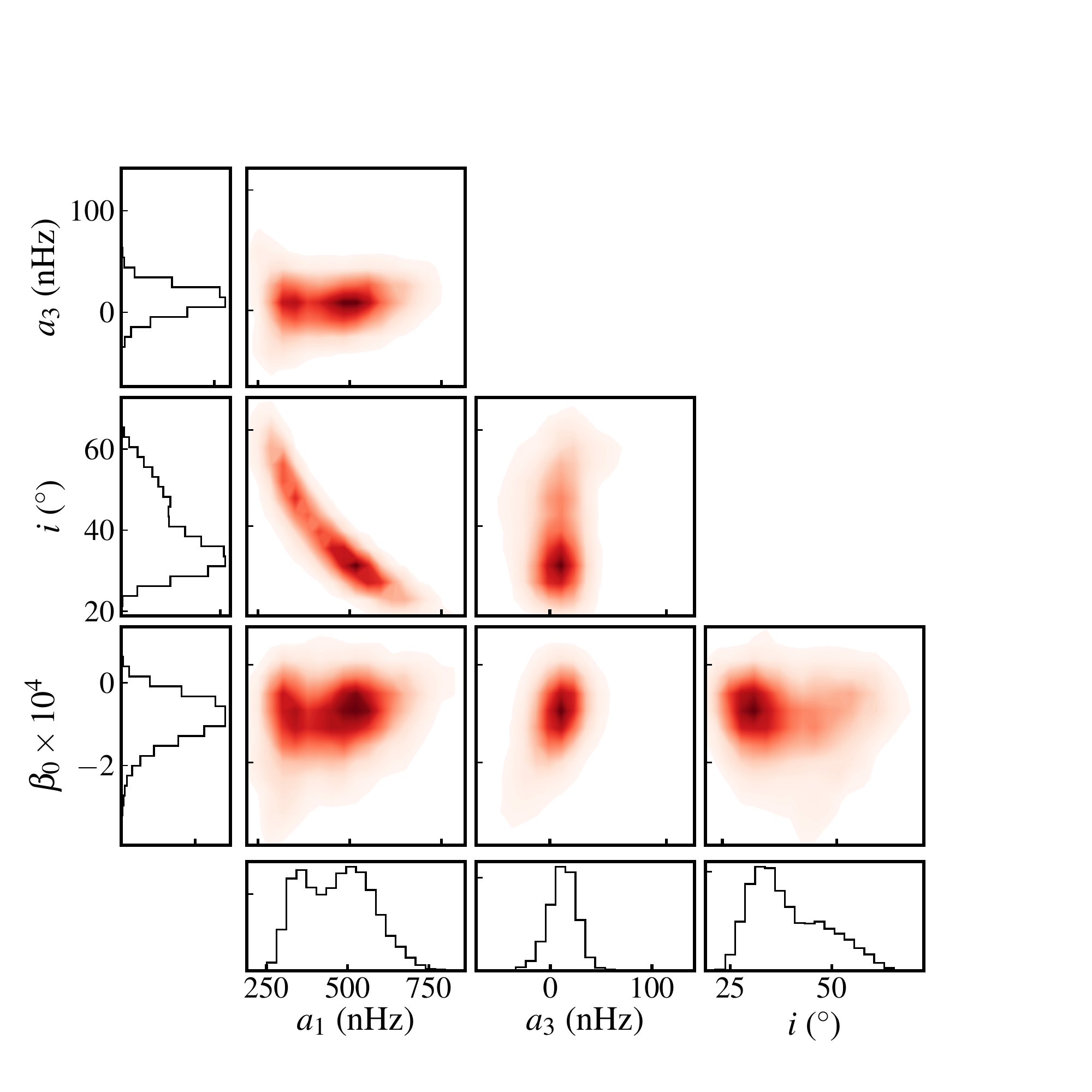}
      \caption{Same as Fig.~\ref{fig:joint-16CygB} for 16~Cyg~B.}
         \label{fig:joint-16CygB}
\end{figure}

The resulting joint marginal densities for $a_1$, $a_3$, $i,$ and $\beta_0$ are displayed in Figs.~\ref{fig:joint-16CygA} and \ref{fig:joint-16CygB}. The parameters seem fairly uncorrelated in general. The only exception worth noting is the obvious trend that can be observed in the joint density of $(a_1,i)$. It is well documented \citep{Gizon13,Nielsen14,Benomar15} that the inferred average rotation rate, which is the main component of the splitting $\delta\nu_{n,l,m}$, increases when the inclination decreases. This is related to the lower visibility of the modes with $m\neq0$ when $i\lesssim 30\degree$, so that only larger splittings can be distinguished at lower angles. Two effects further complicate the fit of multiplets. First, if the mode lifetimes are too short, the splitting and widths of the modes become comparable, producing the blending of the non-degenerate modes. We recall that the mode lifetime $\tau$ is related to the width of the Lorentzian at half-maximum by $\Gamma = 2/\tau$ \citep[see  e.g.  ][]{Chaplin99}. Second, $l=0$ modes may obstruct $l=2$ modes, which are of smaller amplitudes, if the widths of the peaks are too large. These effects will be aggravated when the signal-to-noise ratio becomes low. 

In Fig.~\ref{fig:joint-16CygA} all marginal densities are normal to a good approximation. The inclination of 16~Cyg~A is $58.5^{\circ}\pm6.8^{\circ}$, which confirms the findings of \citet{Davies15}, who obtained a posterior estimate of $56\degree\,^{+6\degree}_{-5\degree}$. As previously discussed, this is an important parameter. The visibility of the modes at such an inclination will be low for the central frequency and higher for $m=\pm1$, as shown in Fig.\ref{fig:spectrum-16CygA}. The estimated value for $a_1$ is $464\pm 43$~nHz. It is in good agreement with the splitting value derived in \citet{Davies15}. After deprojecting, their result is $486.3^{+40}_{-29}$~nHz. The first main result of this study is that we obtain a probability of 86\% for $a_3$ to be strictly positive. Specifying further, we can define a 68.3\% credible interval $a_3 = 11.15\pm10.95$~nHz (given the Gaussian shape of the marginal density, this can be interpreted as a $1\sigma$ interval). Finally, the asphericity parameter is $\beta_0 = 0.1\pm0.1$, that is, it is non-zero at a $1\sigma$ level. The significance of this result is discussed in Sect.~\ref{sect:blabla} in relation to the magnetic properties of the star.

The situation is different for 16 Cyg B, as shown in Fig.~\ref{fig:joint-16CygB}. The probability densities of two parameters, namely $i$ and $a_1$ ,  show a bimodal behaviour. The most likely explanation is a poor constraint on the inclination, which is also reported in \citet{Davies15}. According to the correlation between the inclination and the rotational splitting described above, this in turn impacts the $a_1$ coefficient. The resulting bimodality of many marginal joint PDFs makes interpreting the inversion results difficult. These are discussed in Sect.~\ref{sect:inversion}. We note here that there are some differences with the results found in \citet{Davies15}. We tried to fit the power spectrum of 16~Cyg~B setting $a_3 = 0$~nHz. In that case, we found the same results as they did. Therefore, we interpret the differences between our study and theirs as due to the including $a_3$.

\section{Inversion of the rotation profile}\label{sect:inversion}

The ultimate goal of this study is to provide a map of the rotation rates of 16~Cyg~A and B. The derivation of the probability densities for $a_1$ and $a_3$ allows us to use methods that initially were developed for helioseismology to do so. The entire procedure relies on expanding the splitting according to Eq.~(\ref{eq:splitting}) and the rotation rate, expressed in rad/s, in the form
\begin{equation}\label{eq:rrate}
  \displaystyle
  \Omega(r,\theta) = \sum_{s=0}^S \Omega_{s}(r)W_{s}(\theta),
\end{equation}
with $r$ the stellar radius and $\theta$ the  co-latitude, the system being considered azimuthally symmetric. 

In general, the orthogonality of the functions $\zeta^{(l)}_j(m)$ ensures that only modes with $s\geq j$ contribute to $a_j$ \citep{Brown89}. However, a particular set of functions $W_{s}$ exists such that there is a one-to-one relation between $a_{2j+1}$ and $\Omega_j$ \citep{Ritzwoller91,Schou94,Pijpers97}. 
This is obtained from the relation between the splitting and the rotation-rate components. If the rotational velocity field can be treated as a small perturbation to the hydrostatic equilibrium \citep{LB67}, the antisymmetric part of the frequency splitting can be related to the rotation rate through the linear integral equation
\begin{equation}\label{eq:splitkernel}
  \displaystyle
  \frac{\nu_{n,l,m} - \nu_{n,l,-m}}{2} = \int_0^{\pi}\int_0^{R_{\star}}K_{n,l,m}(r,\theta)\frac{\Omega(r,\theta)}{2\pi}rd\theta dr.
\end{equation}
\begin{figure*}
   \centering
   \includegraphics[width=0.9\textwidth]{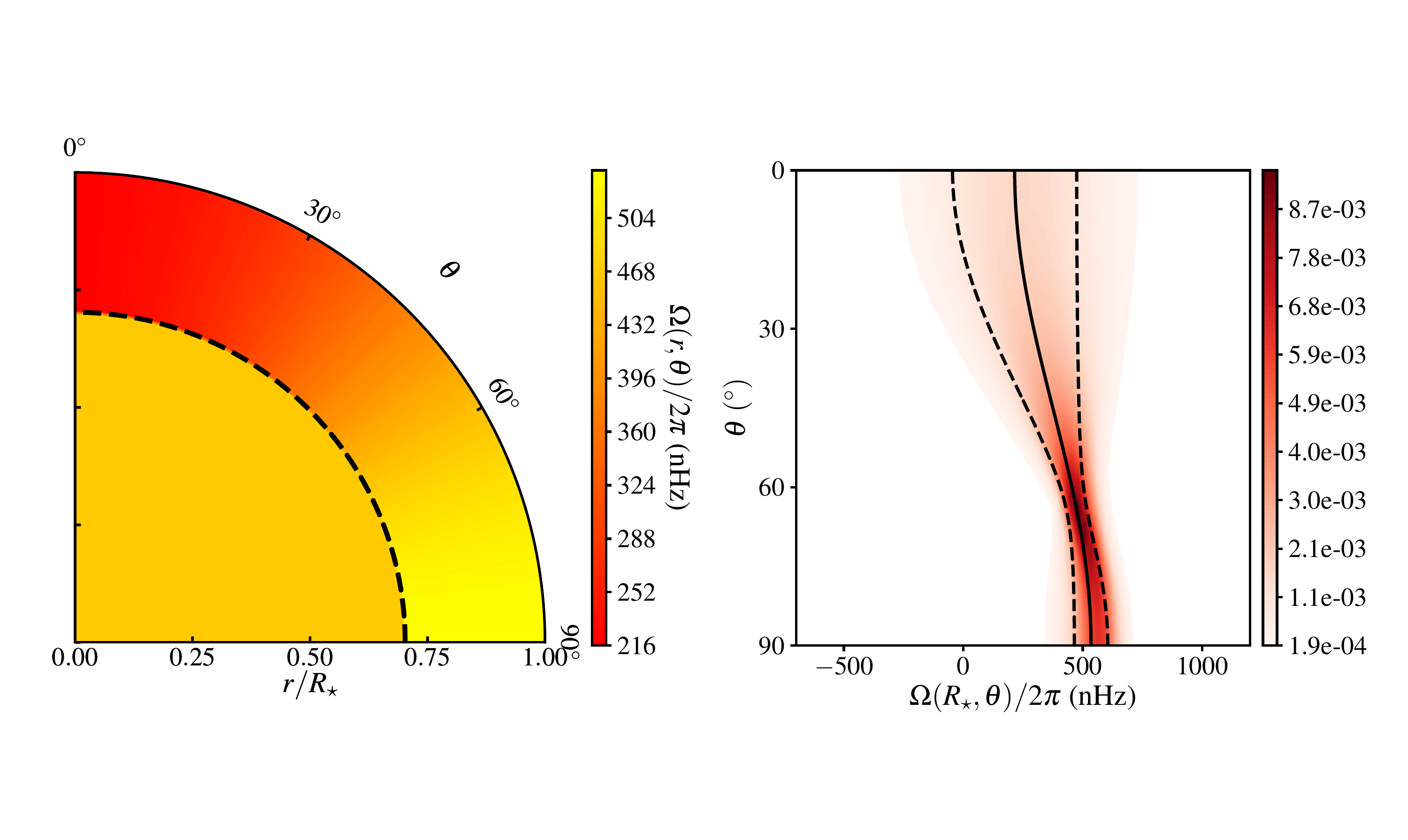}
   \caption{{\bf Left panel}: Rotation-rate profile of 16~Cyg~A corresponding to the PM estimates of $\Omega_0$ and $\Omega_1$. The dashed line marks the bottom of the convective envelope in the assumed stellar model. {\bf Right panel}: Distribution of the surface rotation rate as a function of the stellar co-latitude. The red shade represents probability densities at each latitude. The black line marks the surface rotation model for the PM values of $\Omega_0$ and $\Omega_1$ and corresponds to the map in the right panel at $r=R_{\star}$. The black dashed lines mark the mean surface rotation model plus (right curve) or minus (left curve) the corresponding posterior standard deviation estimated at each latitude.} 
         \label{fig:profile-16CygA}
\end{figure*}

Here $K_{n,l,m}$  is the rotation kernel for the mode defined by $n$, $l,$ and $m$, and expressed as \citep{Hansen77}
\begin{multline}\label{eq:kernel}
\displaystyle
K_{n,l,m}(r,\theta) = \frac{m}{I_{n,l}} \left[ \xi_{n,l}(r)\left[\xi_{n,l}(r) - \frac{2}{L} \eta_{n,l}(r)\right]P_{l}^{m}(x)^2 \right.\\ 
+ \left(\frac{\eta_{n,l}(r)}{L}\right)^2\left[  \left( \frac{dP_{l}^{m}}{dx}\right)^2(1 - x^2) + 2P_{l}^{m}\frac{dP_{l}^{m}}{dx}x \right]\\
 \left.+ \frac{m^2}{1-x^2}P_{l}^{m}(x)^2\right]\rho(r)r\sin\theta.
\end{multline}
We recall the classical notation used above: considering spherical coordinates defined by the basis $(\boldsymbol{e_r},\boldsymbol{e_{\theta}},\boldsymbol{e_{\phi}})$, the total displacement of a fluid element from its equilibrium state is
\begin{equation*}\label{eq:displacement}
\displaystyle
\boldsymbol{\delta r}_{n,l,m}(\boldsymbol{r}) = \xi_{n,l}(r)Y_l^m(\theta,\phi)\boldsymbol{e_r} + \eta_{n,l}(r)\nabla_h Y_l^m(\theta,\phi),
\end{equation*}
with $\boldsymbol{r}$ the position vector and $Y_l^m(\theta,\phi)$ the spherical harmonics. 
We have used the normalised associated Legendre polynomials, $P_{l}^{m}$, and defined $x = \cos\theta$. We also introduced the mode inertia 
\begin{equation}\label{eq:inertia}
\displaystyle
I_{n,l} = \int_0^R \rho(r)r^2[\xi_{n,l}(r)^2 + \eta_{n,l}(r)^2]dr.
\end{equation}

In the following, we assume that the radius-dependent coefficients $\Omega_s(r)$ are piecewise constant functions that can be written explicitly as
\begin{equation}\label{eq:rrate2}
\displaystyle
\Omega(r,\theta)=\begin{cases}
\Omega_0                      &\text{if $r < R_{\text{CZ}}$,}\\
\Omega_0 + \Omega_1W_1(\theta) &\text{if $r \geq R_{\text{CZ}}$.}\\
\end{cases}
\end{equation}
We have used $W_1(\theta) = 1.5(5\cos^2\theta - 1)$ for the first-order basis function \citep{Ritzwoller91,Schou94}.
The kernels were computed from Eq.~(\ref{eq:kernel}) using the output of the code for stellar structure and evolution ASTEC \citep{JCD08a} and the stellar pulsation code {\tt adipls} \citep{JCD08b}. The results, discussed below, are summarised in Table~\ref{tab:rotation}.

It may then be shown \citep{Gizon04} that
\begin{align}
  \displaystyle
  \label{eq:inversion1}2\pi a_1 &= \Omega_0\int_0^{\pi}\int_{0}^{R_{\star}} K_{2,2}(r,\theta)rd\theta dr,\\
\label{eq:inversion12}  &= \Omega_0\Kz ,\\
  \label{eq:inversion2}2\pi a_3 &= \Omega_1\int_0^{\pi}\int_{R_{\mathrm{CZ}}}^{R_{\star}} \frac{[K_{2,2}(r,\theta) - K_{2,1}(r,\theta)]}{5}W_1(\theta)rd\theta dr,\\
  \label{eq:inversion22}&= \Omega_1\Ko,
\end{align}
where $R_{\star}$ and $R_{\mathrm{CZ}}$ are the radius of the star and of the bottom of its convective envelope, respectively. The functions $K_{l,m} = \langle K_{n,l,m}\rangle_n$ are the rotational kernels $K_{n,l,m}$ averaged, with equal weights, over the radial orders.

The derivation of Eq.~(\ref{eq:inversion2}) is straightforward using Eq.~(\ref{eq:splitkernel}) and the expression for $\zeta_3^{(2)}(m)$. To obtain Eq.~(\ref{eq:inversion1}), we assumed that $\delta\nu_{n,1,1} \simeq \delta\nu_{n,2,2}/2$. This is well justified for 16 Cyg A and B. The typical departure between the two splittings is usually below 2\%, which is well below the final uncertainties obtained on $\Omega_0$, of the order of 10\% or more. This assumption allowed us to use $a_1$ as estimated for modes with $l=2$. This could be an advantage since the splittings are easier to measure for higher degree. 

The rotational kernels are thus key quantities to the forward problem, that is, computing the rotational splitting using a theoretical stellar model. They depend in particular on the density of the stratified equilibrium model, $\rho(r)$, and on the radial and horizontal mode displacements $\xi_{n,l}(r)$ and $\eta_{n,l}(r)$. These functions can be obtained by solving the equations for stellar structure, evolution, and pulsation. In order to compute them, a prerequisite is to obtain a realistic stellar model. 
As described in Appendix~\ref{app:model}, the model was obtained by
fitting observed stellar properties, including asteroseismic data.
In this section, we consider the best stellar models we obtained from our simulations. They are defined as those that maximise the posterior density function for the stellar parameters and are given in Table~\ref{tab:params}.

After we computed the stellar models for 16~Cyg~A and B are computed, the rotation kernels were obtained using Eqs.~(\ref{eq:inversion1}) and (\ref{eq:inversion2}). We note that these two relations are sufficient to invert the profile because we only have access to $a_1$ and $a_3$. We tested spectrum models that included $a_5$ (sensitive only to $l=3$ modes) in the truncation of the sum of the right-hand side of Eq.~(\ref{eq:rrate}), but did not detect any significant departure from zero for this coefficient.

\begin{table}
\center
\caption{Differential rotation parameters for 16~Cyg~A and B. For the former, the posterior mean and standard deviation are given in nHz. For the latter, we give the parameters of the three-component Gaussian mixture model used to represent the joint posterior density, as inferred using the EM algorithm. The weights are $p_k$, $k=1,2,3$. The vector ($\Omega_0/2\pi$, $\Omega_1/2\pi$) is the mean vector of each component (in nHz). The coefficients $\sigma_{i,j}$, $i,j=1,2$, of the covariance matrices are given in nHz$^2$. We note that $\sigma_{1,2} = \sigma_{2,1}$.}
\label{tab:rotation}      
\centering                                      
\begin{tabular}{lccc}          
\hline\hline
\\[-2.mm]
\multicolumn{4}{c}{16~Cyg~A}\\
\cmidrule(r){2-4}
\\[-3.mm]
$\Omega_0/2\pi$ & \multicolumn{3}{c}{$471\pm43$}\\
$\Omega_1/2\pi$ & \multicolumn{3}{c}{$-42.7 \pm 41.5$}\\[3.mm]
\multicolumn{4}{c}{16~Cyg~B}\\
\cmidrule(r){2-4}
& $p_1$ = 0.40 & $p_2$ = 0.33 & $p_3$ = 0.27 \\
\cmidrule(r){2-2} \cmidrule(r){3-3} \cmidrule(r){4-4}
$\Omega_0/2\pi$    & 518  & 385  & 618  \\
$\Omega_1/2\pi$    & -47.1  & -43.6  & -68.1  \\
$\sigma_{1,1}$ & 2562 & 1557 & 4157 \\
$\sigma_{2,2}$ & 2029 & 3873 & 1849 \\
$\sigma_{1,2}$ & -108 & 60.1   & -647 \\
\hline
\end{tabular}
\end{table}

\subsection{16 Cyg A}\label{sect:16CygA}

The inversion for the rotation profile can be carried out straightforwardly for 16~Cyg~A. The posterior marginal densities for $a_1$ and $a_3$, $\pi_{a_1\vert\yv}$ and $\pi_{a_3\vert\yv}$, which we plot in the side panels of Fig.~\ref{fig:joint-16CygA}, $\Kz$ and $\Ko$, are known quantities. We can therefore obtain the posterior densities for $\Omega_0$ and $\Omega_1$. We applied Eqs.~(\ref{eq:inversion12}) and (\ref{eq:inversion22}) to the samples obtained from the MCMC simulations $\{ a_1^{(1)},\dots,a_1^{(T)} \}$ and $\{ a_3^{(1)},\dots,a_3^{(T)}\}$, with $T$ the number of realisations in our sample. These are distributed approximately as $\pi_{a_1\vert\yv}$ and $\pi_{a_3\vert\yv}$.
This scaling gives us two new samples $\{ a_1^{(1)}/2\pi\Kz,\dots,a_1^{(T)}/2\pi\Kz \}\sim\pi_{\Omega_0\vert\yv}$ and $\{ a_3^{(1)}/2\pi\Ko,\dots,a_3^{(T)}/2\pi\Ko\}\sim\pi_{\Omega_1\vert\yv}$ (with the symbol {\lq}$\sim${\rq} meaning {\lq}distributed as{\rq}), which we used to approximate the desired marginal posterior probability densities.

\begin{figure*}
   \centering
   \includegraphics[width=0.45\textwidth]{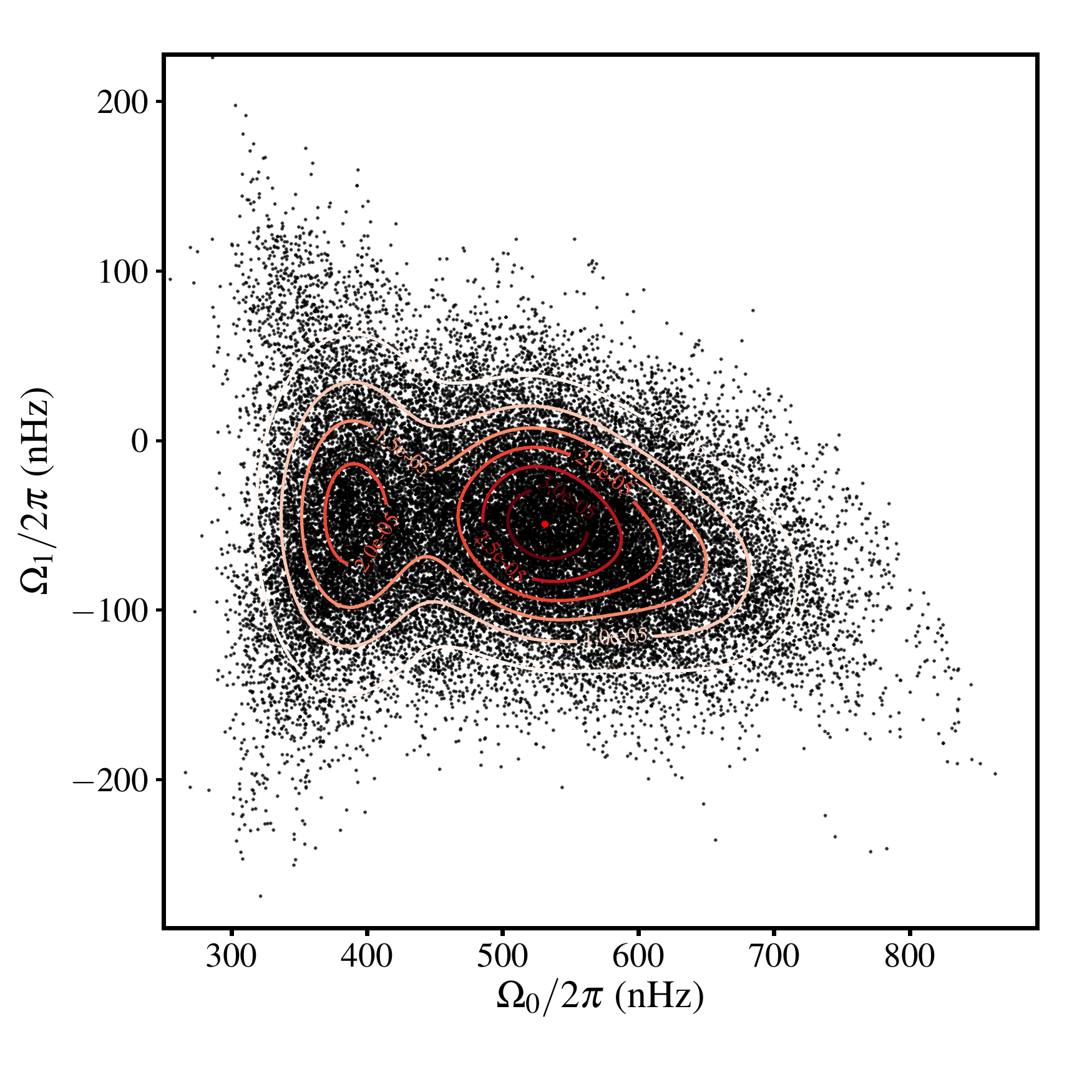}
   \includegraphics[width=0.45\textwidth]{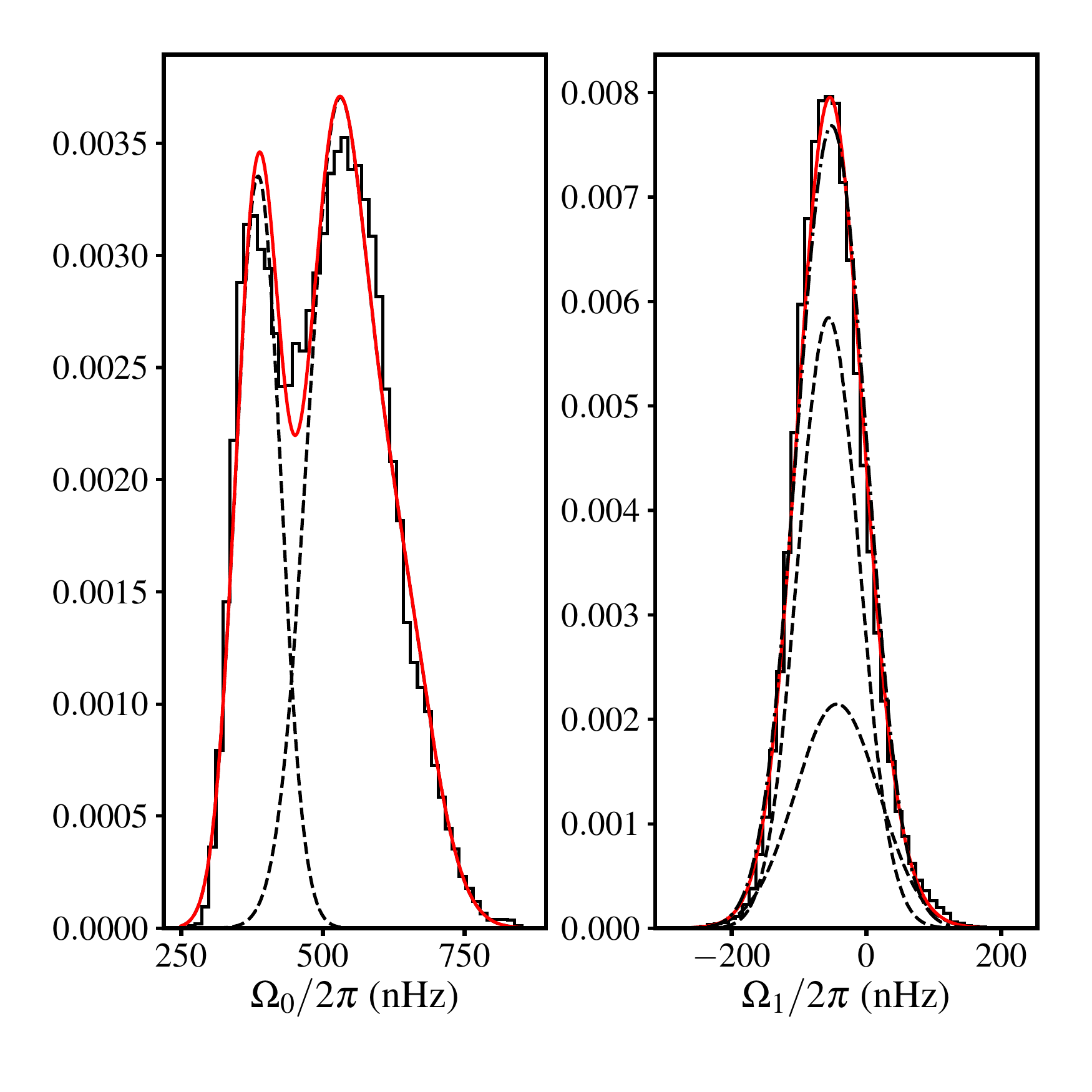}
   \caption{{\bf Left panel}: Joint marginal posterior density for $(\Omega_0,\Omega_1)$ in 16~Cyg~B. The black dots are the MCMC sample. The continuous curves show some iso-probability levels of the corresponding three-component Gaussian mixture model. {\bf Right panel}: Projections on the $\Omega_0$ and $\Omega_1$ axis of the two-dimensional joint probability. The histograms show the MCMC sample and the red lines the mixture model. The dashed lines show contributions of the two main peaks in the joint density (where the main peak combines components 1 and 3, and the secondary peak is component 2; see Table~\ref{tab:rotation}). The dot-dashed line in the $\Omega_1$ projection shows the result using a normal distribution with the mean and variance of the MCMC sample.} 
         \label{fig:joint-a1a3-16CygB}
\end{figure*}

This simple scaling obviously preserves the general structure of the posterior probability densities of the splitting coefficients. The densities for $\Omega_0$ and $\Omega_1$ are thus approximately described by normal densities. We estimated, in the sense of the posterior mean (PM), $\Omega_0/2\pi = 471\pm 43$~nHz and $\Omega_1/2\pi = -42.7 \pm 41.5$~nHz. Here, we used the posterior standard deviation as a 68.3\% credible interval. As expected, the non-null $1\sigma$-level detection of latitudinal differential rotation obtained from $a_3$ translates into the rotation-rate coefficient. As before, the sign of $\Omega_1$ can be assigned a probability, and the probability for it to be negative is still 86\%.

The profile corresponding to the PM of $\Omega_0$ and $\Omega_1$ is given in Fig.~\ref{fig:profile-16CygA} (left panel). It ranges from 534~nHz at the equator to 215~nHz at the pole, that is, a ratio of 2.5, which is significantly higher than that observed for the Sun. The uncertainties on this rotation profile are also shown in Fig.~\ref{fig:profile-16CygA} (right panel). Their behaviour is a good indication of the regions of the stellar surface we can probe efficiently using the current seismic data. The overall structure of the posterior standard deviation seen in Fig.~\ref{fig:profile-16CygA} is due to the functional form retained for $\Omega$. The variance of $\Omega(R_{\star},\theta)$ for a given co-latitude $\theta$ results from the posterior variances of $\Omega_0$ and $\Omega_1$. In Eq.~(\ref{eq:rrate2}) the term depending on $\Omega_0$ does not vary with $\theta$, hence it implies a minimum uncertainty on the rotation rate at any latitude. In contrast, $\Omega_1$ is modulated by $W_1$ , and we have $W_1(\theta = 63.4\degree) = 0$. At this co-latitude, the rotation rate is equal in the radiative and convective envelopes and the variance on the rotation rate is minimal since $\Omega_1$ does not contribute. 

\begin{figure*}
   \centering
   \includegraphics[width=0.9\textwidth]{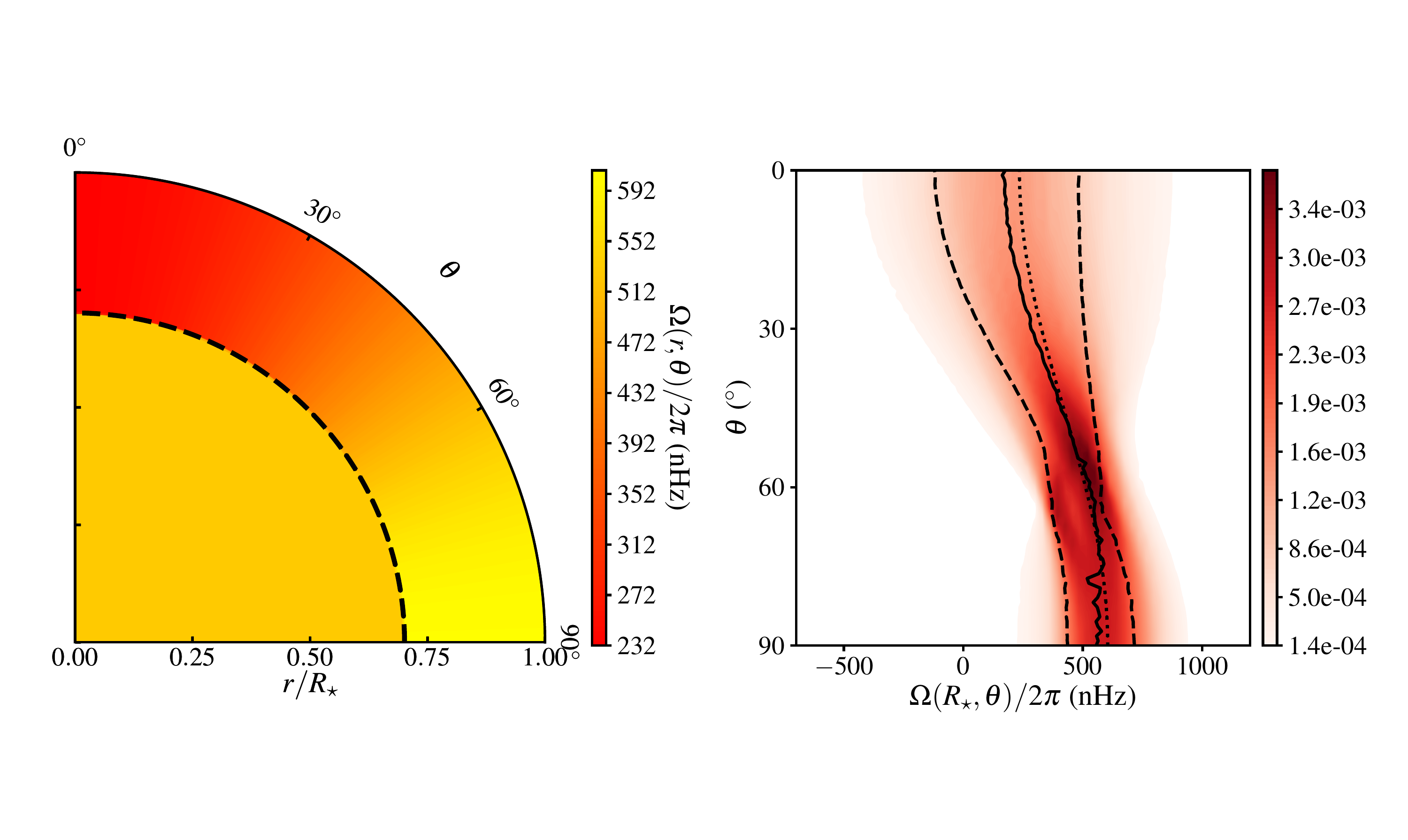}
      \caption{{\bf Left panel}: Rotation profile $\Omega(r,\theta)$ of 16~Cyg~B corresponding to the MAP estimates of $\Omega_0$ and $\Omega_1$. The dashed line marks the bottom of the convective envelope. {\bf Right panel}: Distribution of the surface rotation rate as a function of the stellar co-latitude. The red shade represent probability densities. They have been normalised with respect to the rotational rate at fixed latitude. The black line marks the mode of the corresponding density, obtained from a Gaussian mixture model. The long-dash line shows the associated 68.3\% credible interval. The short-dash line shows the model corresponding to the MAP estimates of $\Omega_0$ and $\Omega_1$.} 
         \label{fig:profile-16CygB}
\end{figure*}

\begin{figure*}
   \centering
   \includegraphics[width=0.9\textwidth]{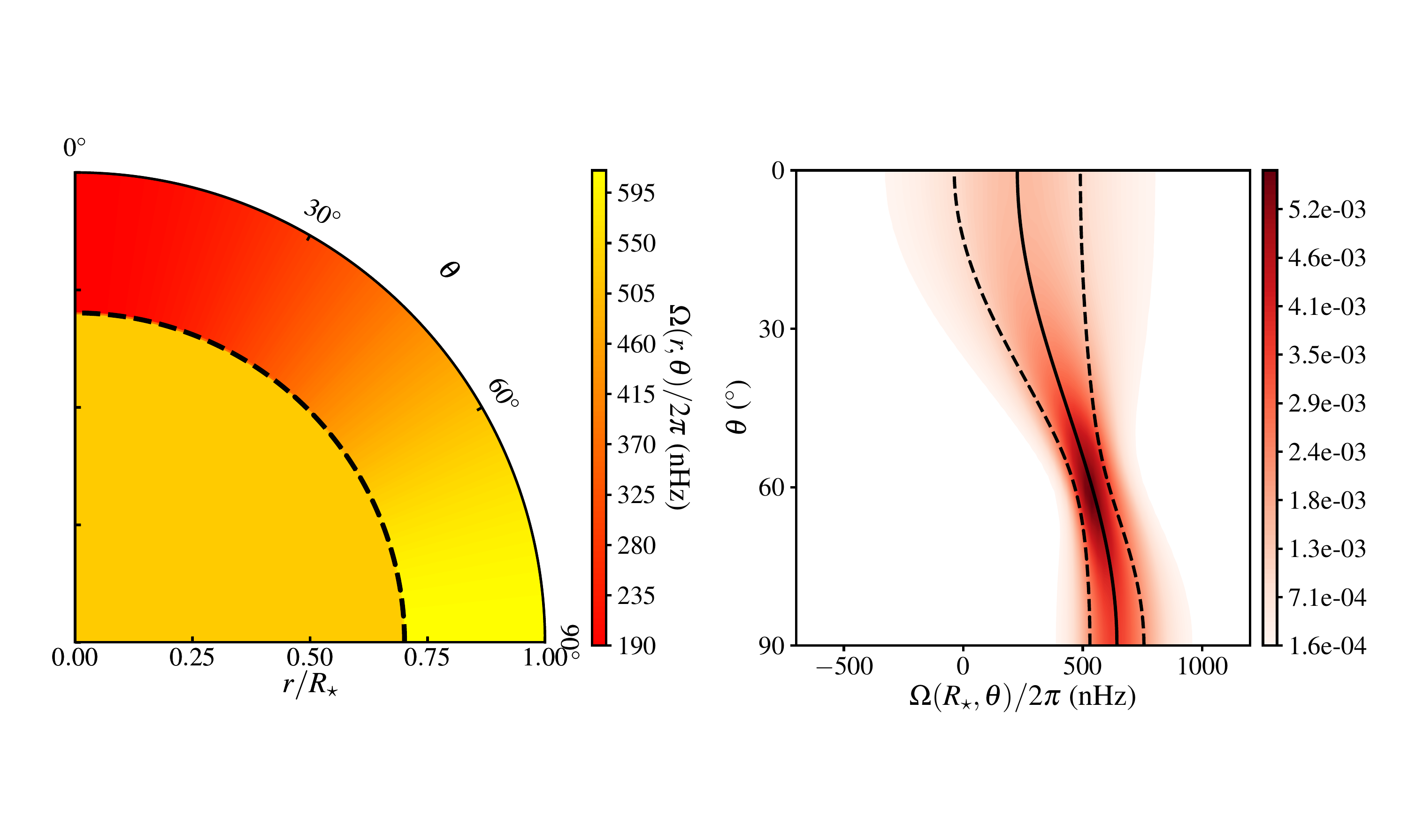}
      \caption{Same as Fig.~\ref{fig:profile-16CygB}, but for the mode corresponding to the two components with the highest mean values in Table~\ref{tab:rotation}.} 
         \label{fig:profile-16CygB-mode}
\end{figure*}
The immediate result is that the higher latitudes are poorly constrained. This is not surprising because even in the solar case, these regions are the most difficult to probe using seismology \citep[see e.g. ][and references therein]{Thompson03}. The uncertainty at co-latitude $0\degree$ is approximately $\pm260$~nHz, which represents an error of $\sim$120\%. The constraint is so poor that some extreme models even show an inversion of their rotation rate between the pole and the equator. Such models, although likely nonphysical, are formally admissible when only Eq.~(\ref{eq:rrate2}) is considered. Thus, the results at high latitudes should not be overinterpreted. The main conclusion that ought to be drawen is that stronger observational constraints are needed to tighten the precision of the inversion near the pole. Such constraints could help to improve our results in two ways. First, by better constraining $\Omega_1$,  large discrepancies of the rotation rate at high latitudes might be prevented. Second, by providing constraints to more accurate models, for instance, expansions of the form of Eq. (\ref{eq:rrate}) at higher orders (typically including $a_5$). We note that using such alternative rotation rates requires kernels that are sensitive to lower co-latitudes 
\citep[see e.g.][]{Lund14}.

Our results are much more constrained near the equator and up to co-latitude $\sim$ $50\degree$. At the equator, the uncertainty on the mean rotation rate is $\pm70$~nHz, amounting to 13\%. It then decreases to 42~nHz at $\sim$65$\degree$, that is, an 8\% uncertainty. The relative statistical error reaches 50\% at roughly 35$\degree$.

\subsection{16 Cyg B}\label{sect:16CygB}

The case of 16~Cyg~B is more subtle and demands greater care. As we have discussed in Sect.~\ref{sect:spectrum}, we observe some correlations in the joint marginal densities of some parameters. At first sight, this does not concern $a_3$. In Fig.~\ref{fig:joint-16CygB} this parameter indeed looks as if it were not correlated to $a_1$ or $i$, and its marginal density $\pi_{a_3\vert\yv}$ appears to be roughly Gaussian. A normal approximation may thus be seen as suitable for modelling this density.

An examination of Eq.~\ref{eq:fdist} gives some hints as to what may cause the observed bimodality. It shows that the frequency spacings between the rotationally split m-components of the $l=2$ modes are $\nu_{n,2, 2} - \nu_{n,2, -2} = 4 ( a_1 + a_3)$ and $\nu_{n,2, 1} - \nu_{n,2, -1} = 2 ( a_1 - a_3 )$. This can lead to degeneracies between $a_1$ and $a_3$ that can only be lifted if the mode blending (typically defined by the ratio $a_1/\Gamma$) and the noise level are not substantial.

When we estimate the first moments of the distribution with our MCMC sample, we obtain $a_3 = 13.89 \pm 13.95$~nHz. Likewise, when we invert $a_3$ using the model discussed above and Eq.~(\ref{eq:inversion2}) to obtain $\Omega_1$, the corresponding sample gives $\Omega_1/2\pi = -51.68 \pm 51.94$~nHz. These values clearly indicate a non-detection of latitudinal differential rotation at a 68.3\% level, that is, zero is included in the $1\sigma$ credible interval when we retain the posterior mean estimator in the Gaussian approximation. This said, the probabilities for $a_3$ to be strictly positive, or conversely, for $\Omega_1$ to be strictly negative, remain high, at $\sim$85\%.

All things being equal, this conclusion remains valid only as long as the normal approximation is accurate enough to describe $\pi_{a_3\vert\yv}$ or $\pi_{\Omega_1\vert\yv}$. In the following, we consider an alternative way to model the joint density $\pi_{\Omega_0,\Omega_1\vert\yv}$ for $(\Omega_0,\Omega_1)$ derived from the joint density $\pi_{a_1,a_3\vert\yv}$. We use a semi-parametric model \citep[see e.g.][]{Bishop95} called Gaussian mixture model. It provides an analytic form that can approximate the density of a random vector $\xv \sim \pi$ and is defined as
\begin{equation}\label{eq:mixt}
  \displaystyle
  \pi(\xv) = \sum_{k=1}^K p_k\mathcal{N}_k(\muv_k,\Sigmav_k).
\end{equation}
Here, $\mathcal{N}_k$ is a multivariate normal density of mean $\muv_k$ and covariance matrix $\Sigmav_k$. The coefficients $p_k$ are such that $\sum_{k}p_k = 1$. All these quantities are parameters that need to be estimated so that they reproduce the observed density satisfactorily, in our case, as approximated by the MCMC sample. We adopted a maximum-likelihood framework to do so. A classical way to obtain estimates of $\muv_k$, $\Sigmav_k$ and $p_k$ is then the expectation-minimisation algorithm \citep[EM,][]{Dempster77}. It is also necessary to fix the number, $K$, of Gaussian components to be used. We proceeded through trial-and-error steps.

Based on these principles, we proceeded in two steps. First, setting $\xv = (\Omega_0, \Omega_1)$ in Eq.~(\ref{eq:mixt}), we modelled the joint density $\pi_{\Omega_0,\Omega_1\vert\yv}$. This can be done straightforwardly using the one-to-one mappings relating $a_1$ and $\Omega_0$, on one hand, and $a_3$ and $\Omega_1$, on the other. Inversion then amounts to independently scaling the components of each vector and preserves the structure of the density. We selected a three-component Gaussian mixture model, which we consider to be the best trade-off between reproducing the main features of the joint probability density and over-fitting. It also has the advantage of being a simple enough model, so that it remains easy to interpret. This approach is still very simple. There exist many subtleties to mixture model fitting, with a vast literature treating them \citep[e.g.][]{FS06}. Our goal here was to show that an improvement of the statistical model could lead us, at the 68.3\% credibility level considered in this study as the reference detection threshold, from a relatively marginal non-detection to a relatively marginal detection. It is clear that from there on, significant improvements require better data. Therefore, we did not pursue more advanced techniques for the mixture modelling.

The results of the mixture modelling are shown in Fig.~\ref{fig:joint-a1a3-16CygB}. The left panel shows the MCMC sample in the $(\Omega_0, \Omega_1)$ plane and the estimated mixture model. The components of the latter can be separated into two groups. Two of them, those with the highest mean values on the $\Omega_0$ axis (components $k=1$ and 3 in Table~\ref{tab:rotation}), account for the peak to the right of the distribution, with a maximum $\Omega_0/2\pi > 500$~nHz. They form the bulk of the density, which can be seen from the fact that the sum of their weights is $0.67$. Two components in the model were necessary to account for the slightly longer tail of the main peak at higher values of $a_1$. The last component, with a weight of $0.33$, represents the mode to the left, peaking at $\Omega_0/2\pi < 400$~nHz. The global mode of the distribution is located at $(\Omega_0/2\pi, \Omega_1/2\pi) = (558~\mathrm{nHz}, -56.6~\mathrm{nHz})$ and is marked by a red dot. The right panel of Fig.~\ref{fig:joint-a1a3-16CygB} shows the projection of $\pi_{\Omega_0,\Omega_1\vert\yv}$ on the $\Omega_0$ and $\Omega_1$ axis. The mixture model accounts for the marginal densities. In particular, it reproduces the two maxima in $\pi_{\Omega_0\vert\yv}$ well. In the case of the marginal density of $\Omega_1$, we also represented the results discussed above and obtained them from a Gaussian approximation. 

We now consider all three components. A first result was obtained by comparing the marginal density for $\Omega_1$ resulting from the projection of the mixture model to the Gaussian approximation. The former is clearly a better approximation. Importantly, when we compute a 68.3\% credible interval relative to its mode, we obtain $\Omega_1/2\pi = -54.45_{-50.06}^{+52.12}$~nHz. This clearly excludes $\Omega_1/2\pi = 0$~nHz. We thus showed that a proper modelling of the density for $\Omega_1$ allows us to obtain a more convincing detection of latitudinal differential rotation. After we established that there is a differential rotation signal in the frequencies of 16 Cyg B, we derived a map of the rotation rate of the star $\Omega(r,\theta)$. Unlike what was done in Sect.~\ref{sect:16CygA}, where we retained the posterior mean (PM) estimates of $\Omega_0$ and $\Omega_1$, we used the MAP estimates of these parameters here. The resulting map is shown in the left panel of Fig.~\ref{fig:profile-16CygB}. The surface rate varies from 604~nHz at the equator to 235~nHz at the pole, for a ratio 2.6, which is only slightly higher that the rate observed for 16~Cyg~A.

The right panel of Fig.~\ref{fig:profile-16CygB} shows the uncertainties on the surface rotation rate. It is apparent in the band of co-latitude 40$\degree$ -- 75$\degree$ that the density is bimodal, reflecting the shape of $\pi_{\Omega_0,\Omega_1\vert\yv}$. Here, we also represent the two estimates of $\Omega(R_{\star},\theta)$ that we obtained: one from the direct modelling of the density $\pi_{\Omega(R_{\star},\theta),\vert\yv}$, the other from the MAP estimate of the parameters $\Omega_0$ and $\Omega_1$. The latter is an approximation to the former. It was convenient to distinguish the two because $\pi_{\Omega(R_{\star},\theta),\vert\yv}$ allowed us to compute a credible interval on the surface rotation rate, while the MAP estimates of $\Omega_0$ and $\Omega_1$ were used to derive the map in the left panel. The approximation is valid since the two solutions never differ by more than $25\%$ of the total width of the credible interval (with a maximum close to the pole) and is, in general, around or below $10\%$ at co-latitudes higher than 15$\degree$. The advantage of the rotation profile based on the MAP estimates of the rotation parameters is that using the parameters given in Table~\ref{tab:rotation}, it can be cast in the close simple analytic form of Eq. (\ref{eq:rrate}) and compared to other studies (see Sect.~\ref{sect:blabla}). The credible interval displayed in Fig.~\ref{fig:profile-16CygB} was not as straightforward to derive as the interval shown in Fig.~\ref{fig:profile-16CygA}, for which we were able to use a Gaussian approximation. In this case we modelled the density of the surface rotation rate at each latitude, $\pi_{\Omega(R_{\star},\theta),\vert\yv}$, using a Gaussian mixture model. The equatorial rotation rate obtained using this estimate is $565_{-129}^{+150}$~nHz. 

The modelling of $\pi_{\Omega_0,\Omega_1\vert\yv}$ allowed us to push the analysis further. It is possible to invert rotation profiles corresponding to the two main peaks found in the density. The less-likely component gives an MAP estimate $\Omega_1/2\pi = -43.6_{-62.5}^{+63.0}$~nHz. The corresponding $a_3$ coefficient is $11.7_{-16.8}^{+16.9}$~nHz, which implies a marginal non-detection at a 68.3\% level if this is the solution (the probability for $\Omega_1$ to be negative remains high, however).

The main peak corresponds to the highest values of $a_1$, with an MAP credible interval $531_{-61}^{+90}$~nHz. The corresponding estimate of the $a_3$ coefficient is $-15.2_{-12.2}^{+12.5}$~nHz. If this turns out to be the solution, then it would correspond to a detection better than the 68.3\% level. We represent the corresponding solution in Fig.~\ref{fig:profile-16CygB-mode} to provide a sense of the results that could be achieved if the data were good enough to constrain the inclination better and, consequently, $a_1$ and $a_3$.

\section{Discussion}\label{sect:blabla}

\subsection{Impact of forward modelling}


Up to this point, our inversion for differential rotation has relied on stellar models obtained from the MAP estimates of the stellar parameters. However, as we show in Appendix~\ref{app:model}, we also estimated uncertainties on these parameters. A legitimate question is thus whether such uncertainties can significantly affect our measurements of latitudinal differential rotation. In particular, the inferred value of $\Omega_1$ could, in theory, be sensitive to the errors on the age and the mixing-length parameter. This could occur through the dependence of the model provided in Eq.~(\ref{eq:rrate2}) on the location of the bottom of the convective zone. The depth of the convective-radiative transition is controlled by the mixing-length parameter, which determines the magnitude of the superadiabatic gradient in the uppermost part of the convection zone and hence the entropy and structure of the adiabatically stratified bulk of the convection zone.
Moreover, on the main sequence, this depth is known to decrease with the stellar age. Therefore, we determined the accuracy of the forward modelling, from stellar parameters to $\Omega_1$, in light of these errors.

In Appendix~\ref{app:model} we obtain approximations to the joint posterior density of the stellar parameters $\thetastv \sim \pi_{\thetastv\vert\Xstv}$, where $\Xstv$ are the observations given in Table~\ref{tab:obs} and $\thetastv$ the stellar parameters. After this, we can also obtain approximations to the densities of a function $h = h(\thetastv)$, that is, $\pi_{h(\thetastv)\vert\Xstv}(h(\thetastv)\vert\Xstv)$. The first step is to formulate a model for the stellar parameters. This can be done using the Bayesian framework we described in Sect.~\ref{sect:spectrum}. 

\begin{figure*}[h!]
   \centering
   \includegraphics[width=.475\textwidth]{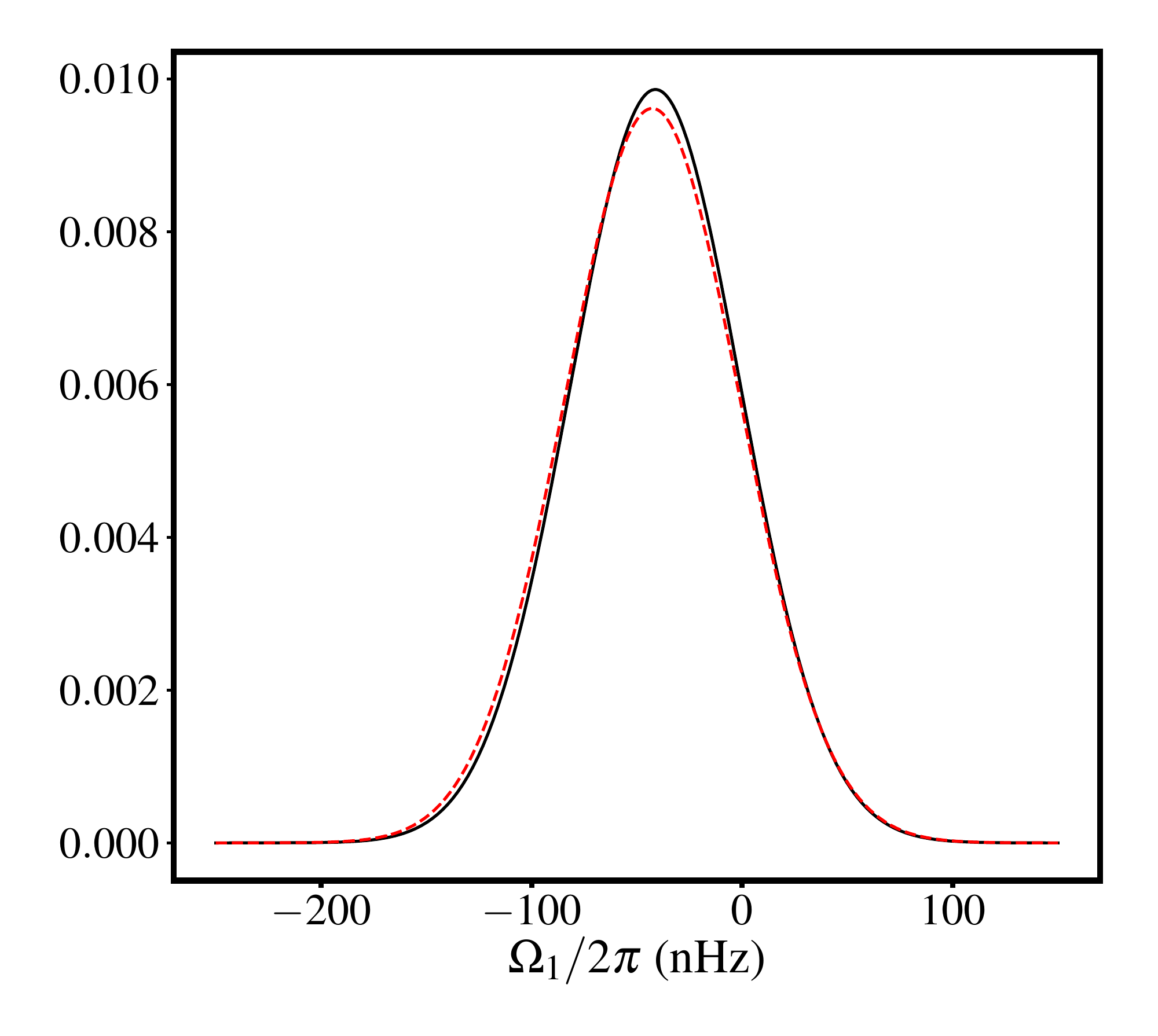}
   \includegraphics[width=.475\textwidth]{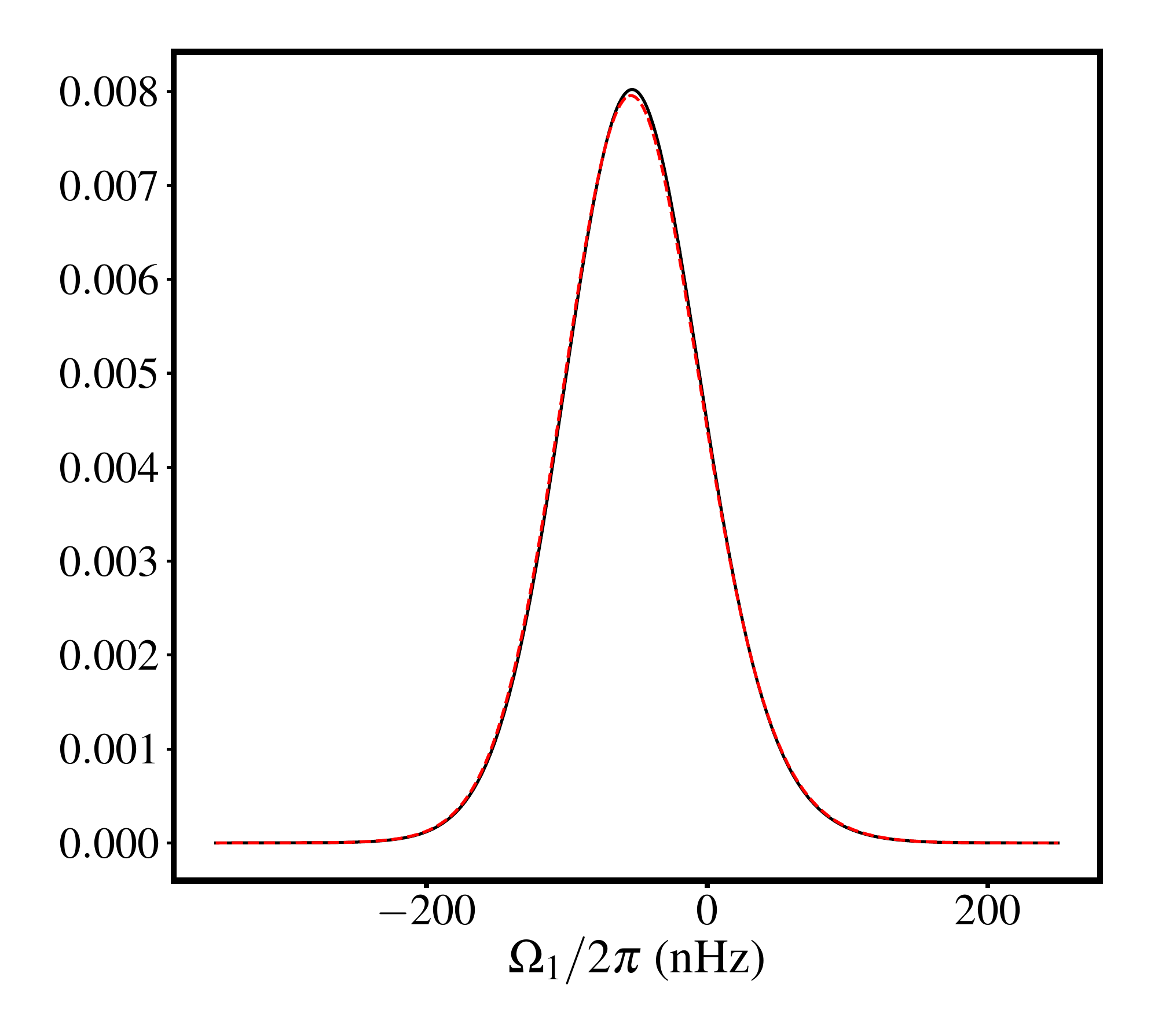}
   \caption{Densities for the differential rotation parameter $\Omega_1$ for 16~Cyg~A (left panel) and B (right panel). The black lines show the densities after marginalisation over the $\Ko$ coefficients given by Eq.~(\ref{eq:pia3}). The red dashed lines show the densities for our best-fit stellar model. }
   \label{fig:16CygAB-omcmc}
\end{figure*}

Table~\ref{tab:params} and Figs.~\ref{fig:joint-16CygA-stellar} and \ref{fig:joint-16CygB-stellar} give for 16~Cyg~A and B the estimates of the stellar parameters in the sense of the PM and the MAP and the marginal two- and one-dimensional probability densities for the stellar parameters, respectively. All the one-dimensional marginal densities are close to Gaussian. Only in the case of $X_0$ is the density slightly truncated as a result of the prior we chose. 


In Fig.~\ref{fig:16CygAB-omcmc} we show the distributions for $\Omega_1$ obtained using the MAP estimates for the stellar parameters, which were used to derive the rotation profiles in Figs.~\ref{fig:profile-16CygA} and \ref{fig:profile-16CygB}. We also display the probability density obtained by taking into account the variability of the scaling coefficient $\Ko$ in Eq.~(\ref{eq:inversion2}) that is induced by the uncertainty on the stellar parameters. In order to obtain this latter density, we assumed that the $a_3$ parameter measured from the acoustic spectrum and the scaling coefficient are statistically independent. This is justified because the effect of rotation on stellar oscillations is treated as a perturbation and we did not take into account the effect of rotation on the stellar structure. In particular, the transport of angular momentum was neglected. Rewriting Eq.~(\ref{eq:inversion2}) as $\Omega_1 = 2\pi a_3/\Ko$, we can derive the probability density $\pi_{\Omega_1\vert\Xstv,\yv}$ for $\Omega_1\vert\Xstv,\yv$ using a standard relation from probability theory that gives the density of a ratio of random variables,
\begin{equation}
  \displaystyle
  \label{eq:densratio}
  \begin{aligned}
    \pi_{\Omega_1\vert\Xstv,\yv}(\Omega_1\vert\Xstv,\yv) =& \int \pi_{a_3\vert\Xstv,\yv}\left(\frac{\Omega_1\mathcal{K}_1}{2\pi}\middle\vert \Xstv,\yv\right)\times\\
    &\pi_{\Ko\vert\Xstv}(\Ko\mid\Xstv)\lvert\Ko\rvert d\Ko, \\
  \approx& \sum_i \pi_{a_3\vert\yv}\left(\frac{\Omega_1\mathcal{K}_{1,i}}{2\pi}\middle\vert\Xstv,\yv\right)\lvert\mathcal{K}_{1,i}\rvert.
  \end{aligned}
\end{equation}
The density $\pi_{\Ko}$ was obtained from the MCMC simulations described above. We note that for any value of the stellar parameters we can compute new oscillation kernels and the corresponding values for the integral $\Ko$, thus obtaining an approximation to the density $\pi_{\Ko}$. The second line in Eq.~(\ref{eq:densratio}) is the approximation of the preceding integral using this MCMC sample. The sum is taken over all realisations. In order to compute this term, we need to know the density for $a_3\vert\yv$. This was done in a similar fashion as for $\Omega_1$. We modelled the joint density for $(a_1,a_3)$ using a three-component mixture model, each of them being bivariate Gaussian densities $\mathcal{N}(\boldsymbol{\zeta}_k,\boldsymbol{\Lambda}_k)$ associated with weights $q_k$, $k = 1,\dots,3$. This model can be marginalised analytically over $a_1$
\begin{equation}\label{eq:pia3}
  \pi_{a_3\vert\yv}(a_3\vert\yv) = 
  \sum_{k=1}^3 q_k\frac{\exp \left[ -\frac{1}{2}\left(\lambda_{2,k} - \frac{\lambda_{12,k}^2}{\lambda_{1,k}} (a_3 - \zeta_{2,k})^2\right)  \right]}{\sqrt{2\pi \lambda_{1,k}}\vert\boldsymbol{\Lambda}_k\vert^{-1/2}},
\end{equation}
with $\zeta_{2,k}$ the second coefficient of $\boldsymbol{\zeta}_k$ and $\lambda_{1,k}$, $\lambda_{2,k}$, and $\lambda_{12,k}$ the coefficients of the co-variance matrix $\boldsymbol{\Lambda}_k$. 

In Fig.~\ref{fig:16CygAB-omcmc} we show that the resulting densities are extremely close to those obtained using the MAP estimates. At any rate, this does not change our conclusions about differential rotation for 16~Cyg~A and B. We can safely assess that differential-rotation measurements depend only weakly on the exact value of the stellar parameters. This is expected to hold as long as rotation can be treated as a perturbation. It is also important that the quality of the seismic data on 16~Cyg~A and B offers a very good precision on the age and mixing-length parameters; this gives us an indication of the range, in the parameter space, over which our results can be regarded as robust. It remains to be understood at which level of precision this breaks down. 

Another difficulty with forward modelling is the lack of information on the terms of higher order in the expansion of $\Omega(r,\theta)$. Potentially, these might counteract the effects of the leading non-constant term in $\cos^2\theta$. This is difficult to assess, however, since we were unable to measure $a_5$. Furthermore, adopting a slightly different decomposition $\Omega(r,\theta) = \sum \Omega_s^{\star}(r)\cos^{2s}\theta$ \citep{Ritzwoller91,Schou94},  $\Omega^{\star}_1$ and $\Omega^{\star}_2$ might be constrained using $a_3$. Of course, we loose in the process the one-to-one relation between  $a_j$ and $\Omega^{\star}_s$. The meagre information obtained from spectrum fitting makes it difficult to properly estimate these parameters. So far, the best argument in favour of the  preservation of latitudinal differential rotation when higher-order terms are included is the extrapolation from the solar case. We know that the term in $\cos^4\theta$ in the above expression also decreases the rotation rate as the co-latitude decreases. A common expression for the solar surface rotation rate in the convective zone is $\Omega_{\odot}(R_{\odot},\theta)/2\pi = 454 - 55\cos^2\theta - 76\cos^4\theta$ \citep{Gizon04}. At the pole, the last term of the right-hand side contributes to $\sim$60\% of the equator-to-pole braking. The missing information on higher-order terms is thus responsible for the poor constraints at high latitudes. As discussed in Sect.~\ref{sect:16CygA}, the form used to describe the rotation rate implies a minimum variance at $\theta = 63.4\degree$. This is no longer the case when we introduce a term in $\cos^4\theta$. In the Sun, this latter dominates differential rotation at co-latitudes $\lesssim 32\degree$, and hence the variance of the surface rotation rate in corresponding proportions. We postulate that the missing information on higher-order terms is responsible for the poor constraining of the rotation rate at high latitudes, $\gtrsim 40\degree$, that we see in Figs.~\ref{fig:profile-16CygA}, \ref{fig:profile-16CygB}, and \ref{fig:profile-16CygB-mode}. A possible solution to this problem is to obtain longer time series, potentially involving several $l=3$ modes, which would allow measuring the $a_5$ coefficients. In that case, there would be a one-to-one relation between the former and the $\Omega_2$ coefficients. These weight the function $W_2(\theta)$ in which the $\cos^4\theta$ terms appear.

We also disregarded the effect of subsurface flows on the rotation rate. It is well established in the solar case that it increases rapidly immediately below the surface \citep{Deubner79}. For regions located in the co-latitude range $60\degree - 90\degree$, the angular velocity gradient remains approximately constant. The layer at $r = 0.97R_{\odot}$ rotates $\sim$3\% faster than the surface \citep{Corbard02}. At lower co-latitudes, the gradient decreases in magnitude as a function of $\theta$ and even becomes positive below $\sim$35$\degree$. This may bias the results presented here. The modes we used to infer $a_3$ are not the most sensitive to these subsurface layers, and the values derived here for latitudinal differential rotation may be more representative of the bulk of the convective zone rather than the surface itself, or the regions immediately below. The subsurface shear layer thus remains to be properly taken into account, as discussed for instance in \citet{Lund14}, in which the change in the rotational rate is uniformly modelled using a latitude-independent gradient. 

\subsection{Other differential rotation measurements}

To our knowledge, no previous detection of differential rotation in 16~Cyg~A or B has been reported so far. It is noteworthy that these stars are included in the BCool snapshot program \citep{Marsden14}. This survey aims at detecting the average longitudinal component of stellar magnetic fields \citep{Semel89,Landi92}. Stars with conclusive detection then undergo further modelling of their magnetic topology. 
As a by-product, this provides an estimate of latitudinal differential rotation. The model used to reproduce the observations is a parametrisation of the magnetic field \citep{Hussain01} whose output is then transformed to reproduce the Zeeman profile. The latter is deconvolved from the observed V Stokes profile \citep{Donati97b,AsensioRamos16}. The mapping from the magnetic field to the one-dimensional Zeeman profile involves a convolution of the theoretical spectrum by the rotation profile of the star, also described here by Eq.~(\ref{eq:rrate2}). The differential rotation parameters are obtained, alongside those describing the magnetic field, using least-square minimisation. Unfortunately, the variations in the V Stokes spectra of 16 Cyg A and B could not be attributed with confidence to the magnetic field, which so far implies that their magnetic activity is not strong enough to allow a proper characterisation of the rotational profile.

\begin{figure}
   \centering
   \includegraphics[width=0.5\textwidth]{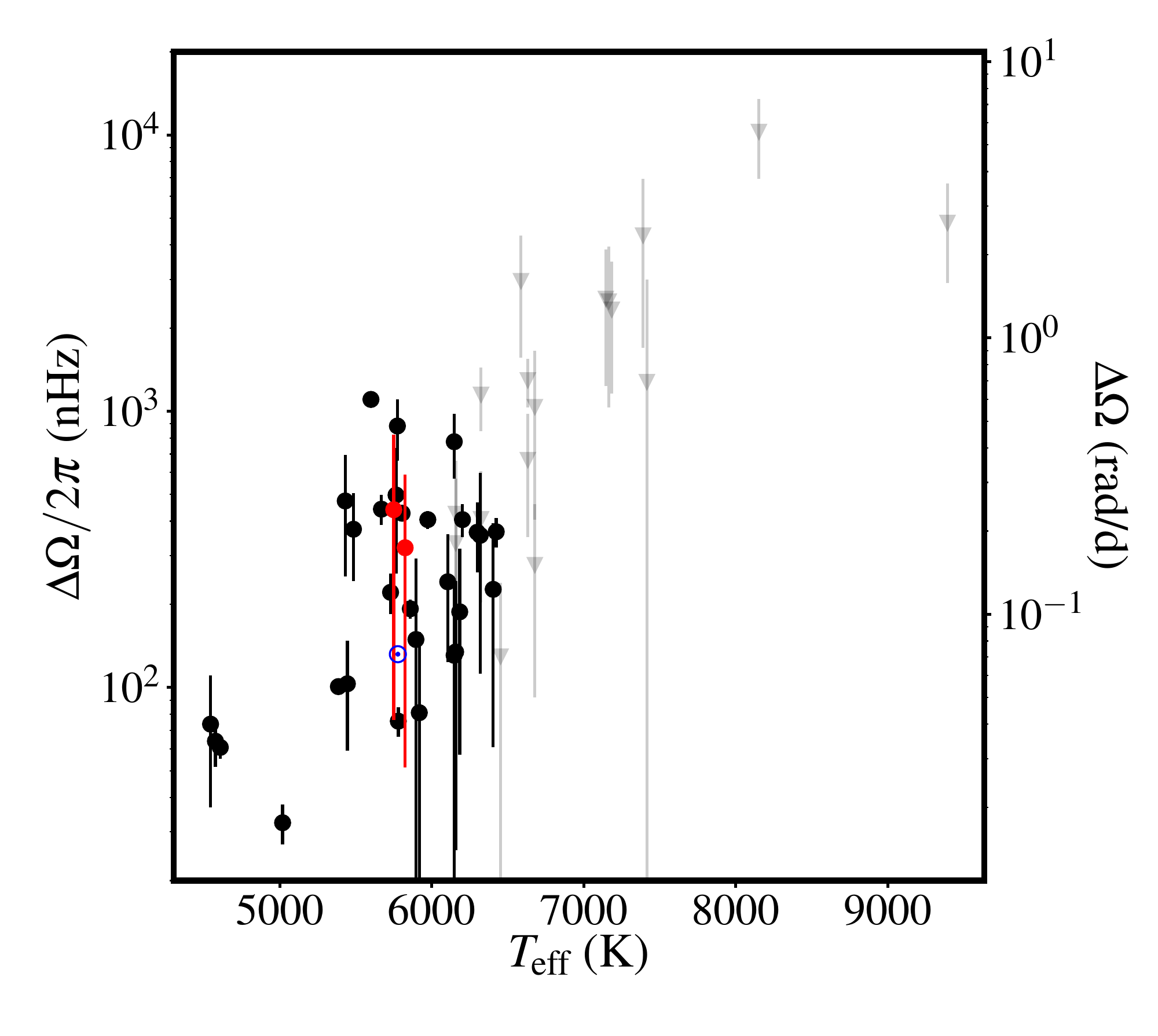}
   \caption{Measured latitudinal differential rotation as a function of the effective temperature for 16~Cyg~A and B (in red) and stars observed using spectropolarimetry, photometric transit, and asteroseismology. The Sun is represented by the blue $\odot$ symbol. The grey triangles represent the upper bound provided by \citet{AvE12}.}   
\label{fig:teff-drot}
\end{figure}

Latitudinal differential rotation is usually quantified using either $\Delta\Omega = \Omega - \Omega(R_{\star},0\degree)$ or the so-called shear parameter $\Delta\Omega/\Omega$, where we set $\Omega = \Omega(R_{\star},90\degree)$, the equatorial rotation rate. In the case of 16 Cyg A and B, we measured $\Delta\Omega/2\pi = 320\pm269$~nHz and $440^{+363}_{-383}$~nHz, respectively, and  $\Delta\Omega/\Omega = 0.65^{+0.47}_{-0.50}$ and $0.76^{+0.55}_{-0.58}$. These limits correspond, as usual, to 68.3\% credible intervals. We note also that the probabilities of the shear parameter to be positive are 85\% and 86\% for 16 Cyg A and B.

We can gain some perspective by comparing these results to the other measurements of latitudinal differential rotation provided by \citet{Benomar18} and to other measurements obtained using spectroscopy. The asteroseismic measurements of these quantities are of a similar order of magnitude as those found for 16 Cyg A and B. However, with the exception of KIC~10963065, all the estimated shear parameters are higher. The most extreme case is KIC~9025370, for which the shear parameter is $\sim$4 times higher than in 16 Cyg A. This may be the reflection of a trend of deceasing magnitude for differential rotation with age. However, several factors may be at work here. As discussed below, the effective temperature (see below) is also of importance. More precise, statistical statements on the full sample of star with asteroseismically-measured latitudinal differential rotation are beyond the scope of this paper. The correlation between differential rotation and other stellar parameters will be considered in future studies. 


Some other measurements of latitudinal differential rotation have also been obtained using observational techniques other than asteroseismology. In the following we focus on results obtained using spectroscopy. Even though claims of latitudinal differential rotation detection have been made using photometry \citep{Reinhold13b,Lanza14}, it was pointed out by \citet{Aigrain15} that the methods considered might not be entirely reliable. Therefore we do not consider them here. The only notable exception is the planet-hosting star Kepler-17 \citep{Valio17}, for which the particular orbital configuration of the planetary system allows a precise measurement of latitudinal differential rotation. However, we recall that the method employed in this study might not lead to many detections in the future. However, even though these investigations used \emph{Kepler} data, these data were not analyzed using asteroseismic techniques.

In the case of spectroscopy, Doppler imaging and (the closely related) Zeeman Doppler imaging have been important providers of latitudinal differential rotation estimates. \citet{Vogt87} initially stated that Doppler imaging requires fast rotators so that it is the dominant mechanism that produces spectral line broadening. The same remark applies to Zeeman Doppler imaging, which can be seen, crudely, as a transposition of Doppler imaging into V Stokes profiles \citep{Semel89,Brown91}. \citet{Petit02} have shown that the method could be applied to moderate rotators to obtain latitudinal differential rotation measurements. In both cases, the methods require stellar spots that modulate the observed spectrum.

Compared to the estimates given in Sect.~\ref{sect:inversion}, Zeeman Doppler imaging often leads to smaller errors on the parameters of the rotation profile. An explanation is that the spot configurations encountered on the observed stars often imply surface tracers distributed on a wide range of latitudes, including, potentially, near the pole. Such observations may therefore constrain the rotation-rate profile over the entire stellar surface, while asteroseismic inversions are controlled by the sensitivity of the observed modes to the regions of the stellar interior that form their resonant cavities. Spectropolarimetric detections are always at levels $\gtrsim 1.5\sigma$ (normal densities are assumed for the uncertainties on the differential rotation parameters), and sometimes better than $40\sigma$.


It  has been suggested by \citet{Barnes05a} that a relation between the effective temperatures of stars and the magnitude of their latitudinal differential rotation exists, giving the power law $\Delta\Omega \propto T_{\mathrm{teff}}^{8.92\pm0.31}$. A theoretical explanation of this trend has been advanced by \citet{Kitchatinov12}, although not all behaviour could always be accounted for \citep{Kuker11}.  In Fig.~\ref{fig:teff-drot} we show a plot similar to Fig.~2 in \citet{Barnes05a}, in which we display the values of the latitudinal differential rotation for 16 Cyg A and B alongside those obtained by \citet{Benomar18}, and others from Zeeman Doppler imaging or spectrographic measurements. For most cases we used the values published in this study for the effective temperature, except for HD 197890, for which we used the value of \citet{Casagrande11}. When several measurements for $\Delta\Omega$ existed, we used a weighted mean and computed the corresponding standard deviation assuming Gaussian errors. We note that by doing so, we consider that  the measurements are realisations of a random process, and we might be disregarding a temporal dependence of these variations that would correlate the measured values. We also added some stars observed later using spectropolarimetry or photometry: Kepler-17 \citep{Valio17}, 61~Cyg~A \citep{BoroSaikia16}, HN~Peg \citep{BoroSaikia15}, $\xi$~Boo \citep{Morgenthaler12}, HD~35296 and HD~29615 \citep{Waite17}, EK~Dra \citep{Waite17}, HD 141943 \citep{Marsden11}, HD10650 \citep{Waite11}, and $\tau$~Boo \citep{Donati08,Fares09}. The inclusion of 16 Cyg A and B does not seem to contradict the law described above. The use of weighted means seems to modify this result more than our new measurements. \citet{Barnes05a} used all the individual measurements to fit the data. 

Alongside these differential-rotation measurements, we also display those given by \citet{AvE12}. They were obtained for A and F stars, which are relatively fast rotators. However, these values have to be considered with caution because they are only upper limits. 

\subsection{Asymmetries in the power spectrum}

\begin{figure}
   \centering
   \includegraphics[width=0.45\textwidth]{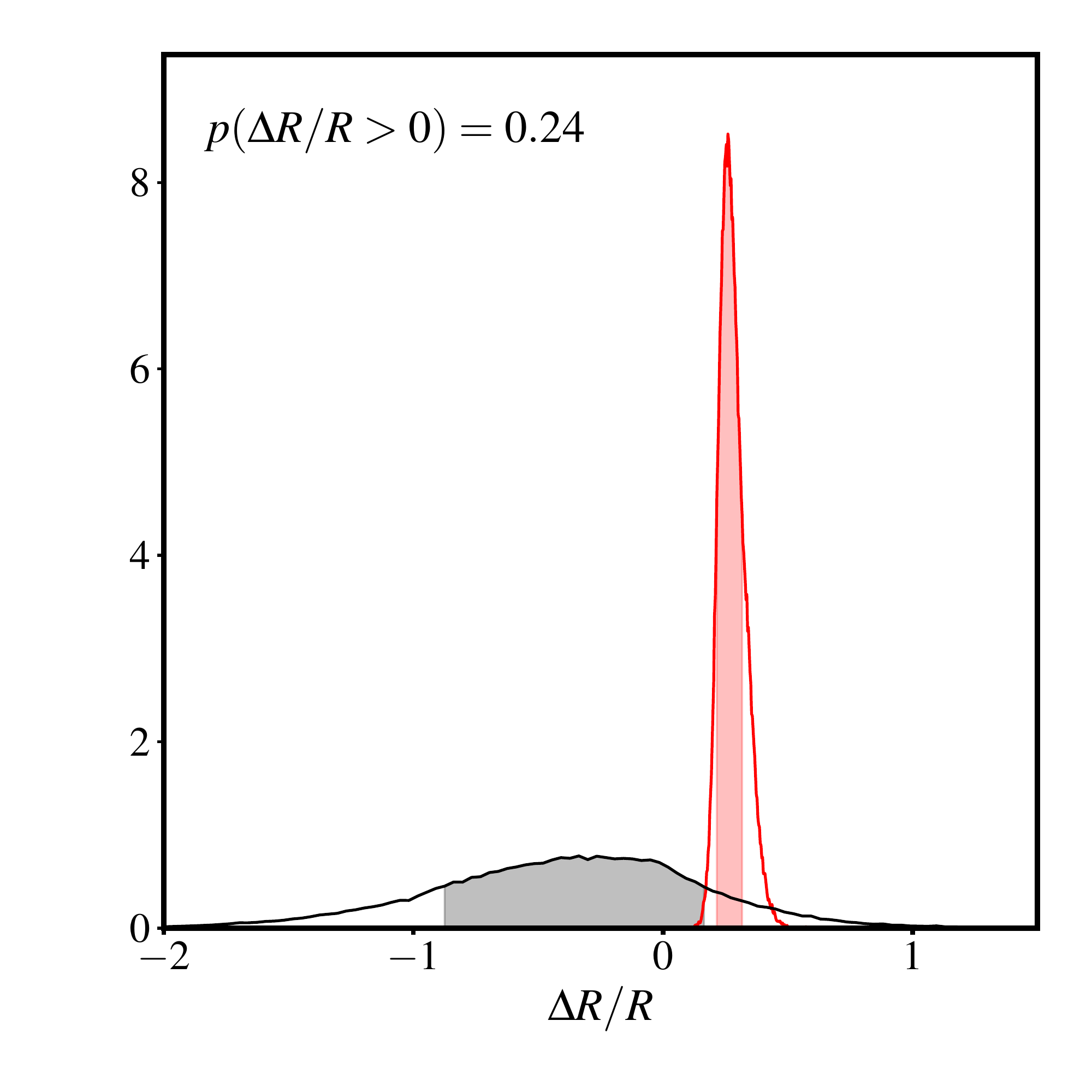}
   \includegraphics[width=0.45\textwidth]{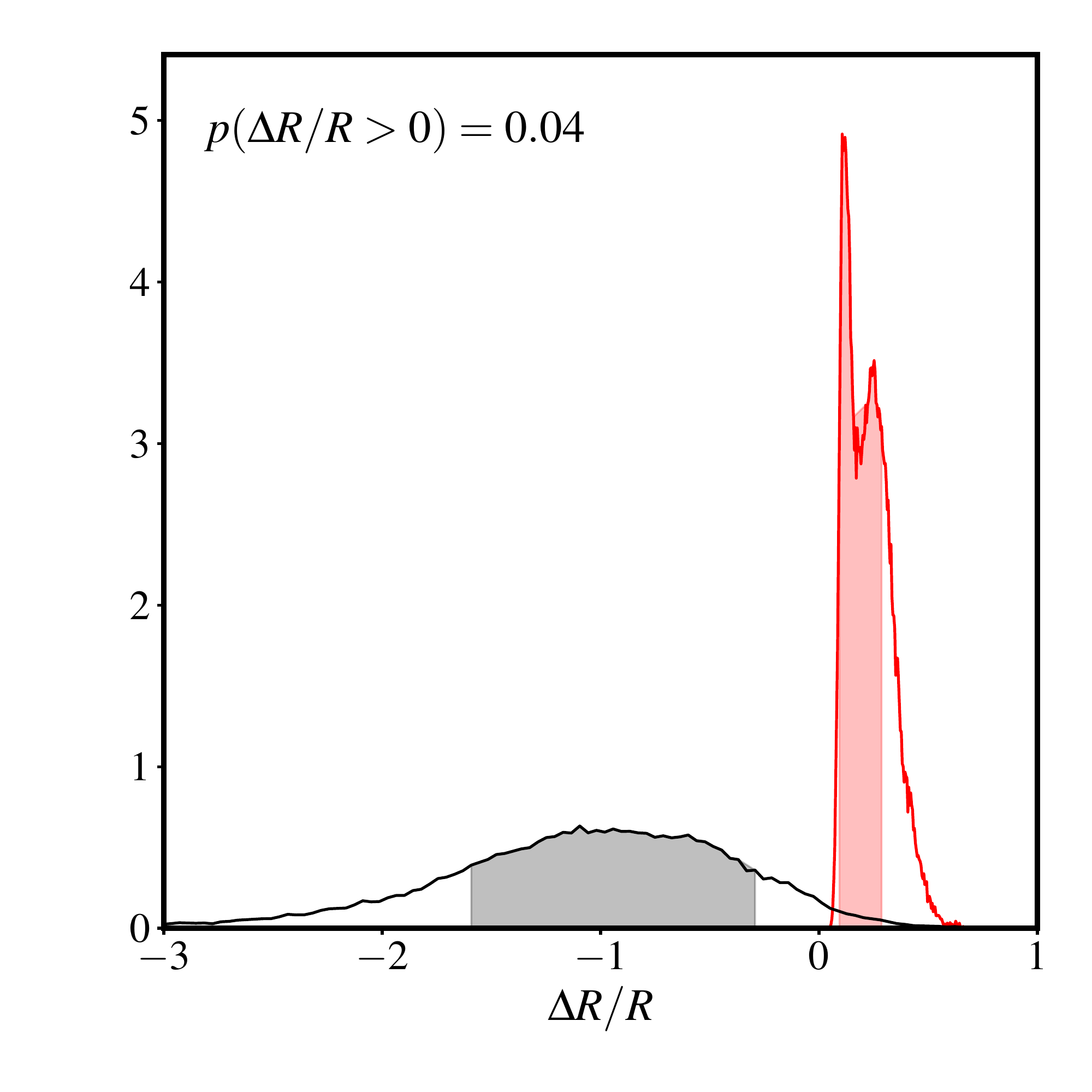}
   \caption{Probability densities for the asphericities of 16~Cyg~A (upper panel) and 16~Cyg~B (lower panel). The red line shows the density resulting from a purely centrifugal force computed from $a_1$. The black line shows the measured effective asphericity $\Delta R/R\vert_{\mathrm{eff}}$. The probability that the star is oblate ($\Delta R/R > 0$) is given. The shaded areas mark the 68.3\% credible intervals of the corresponding distributions.}   
\label{fig:asphericity}
\end{figure} 

The last point we wish to discuss is the deformation of the star by rotation and its effect on frequencies. Because stars are spheroids in rotation, it is evident that the centrifugal force may play a major role in determining the shape of the star.



Devising a method to detect the asphericity of a star might therefore give clues about the ongoing mechanisms in the stellar interior and at the surface. Interestingly, while the centrifugal force causes the star to become oblate, this is not necessarily the case for other forces. Typically, a toroidal magnetic field counteracts the effect of the centrifugal force such that the star may become prolate \citep{Chandra53,Wentzel61}. 

The term $\beta_{n,l,m}$ defined in Eq.~(\ref{eq:beta}) is only due to the centrifugal force, and so we can write
\begin{align}
  \displaystyle
        \beta_{n,l,m} &= \frac{4 \pi}{3} \frac{Q_{lm}}{G \rho_\star}\,\nu_{n,l}\,\Omega^2(r, \theta=0) \\
                        &\approx \frac{4 \pi}{3 G \rho_\odot} \frac{\Delta\nu^2_\odot}{\Delta\nu^2}\,Q_{lm}\ \nu \,a^2_1 = \eta_0 \,Q_{lm}\ \nu_{n,l}\, a^2_1,
                \label{eq:beta:centforce}
\end{align}
where $G$ is the gravitational constant and $\rho_\star$ is the average density of the star, approximated by the measure of the frequency spacing $\Delta\nu$ of the pressure modes, $\rho_\star \approx \rho_\odot \,(\Delta\nu/\Delta\nu_\odot)^2$. Here we also assumed that the $a_1$ coefficient is representative of the the equatorial rotation. In the light of the result from the seismic inversion, this is indeed justified.

The actual pole-to-equator distortion of the star is then \citep{Gizon16}
\begin{equation} \label{eq:asph_cenforce}
  \displaystyle
        \frac{\Delta R_c}{R} = \frac{3}{8 \pi} \, \eta_0 \, a^2_1,
\end{equation}
with $\Delta R_c = R_{\mathrm{eq}} - R_{\mathrm{pole}}$. Here $R_{\mathrm{eq}}$ is the equatorial radius and $R_{\mathrm{pole}}$ the polar radius of the star.

To measure an asphericity that is not only due to the centrifugal force, the general functional form $\beta_{n,l,m} = \beta_0 \,Q_{lm}\,\nu_{n,l}$ could be used. This relates to an effective asphericity coefficient $\Delta R/R$ by
\begin{equation} \label{eq:asph_all}
  \displaystyle
        \left.\frac{\Delta R}{R}\right\vert_{\mathrm{eff}} = \frac{3}{8 \pi} \beta_0.\end{equation}
This choice allows us to straightforwardly compare the case of a pure centrifugal force with cases with additional distorting forces.

 In Fig.~\ref{fig:asphericity} we compare the effective ashericities and those computed from $a_1$ for 16 Cyg A and 16 Cyg B. They were determined using Eqs.~(\ref{eq:asph_cenforce}) and (\ref{eq:asph_all}). In the case of 16~Cyg~B, although we cannot reject the possibility that the star is oblate ($24.4\%$ probability), it is striking that the measured asphericity is only marginally consistent with the case of a pure centrifugal force. The discrepancy is even more significant when considering 16 Cyg B, as the probability of being oblate is only $3.9\%$. One possible interpretation is that 16 Cyg A and B have a consequent (measurable) toroidal (equatorial) field that dampens equatorial waves such that they travel in a prolate cavity.
 
\section{Conclusions}


We have reported the detection of differential rotation for the two good solar analogues 16 Cyg A and B. We followed \citet{Benomar18}, where latitudinal differential rotation was detected in a sample of stars that either rotate faster than the Sun and/or with much higher differential rotation. In this case, the inferred values for the stellar rotation rate and for differential rotation are consistent with the solar regime.

We have described a way to model the latitudinal rotational splitting of 16 Cyg A and B by taking into account the impact of differential rotation. Using a Bayesian setting, we were able to state that the $a_3$ coefficients in these stars have a probability $\gtrsim 85\%$ of being strictly positive. Using an expansion basis for which a one-to-one relation exists between the coefficients for the rotational splitting and rotation rate, we translated this result into probabilistic statements on $\Omega_1$, the coefficient quantifying the amount of latitudinal differential rotation in the convective zone. Importantly, it has the same $\gtrsim85\%$ probability of being strictly negative for both stars. This indicates that it is very likely that the azimuthal component of the flow in the convective zone undergoes an equator-to-pole braking. We also provided summary statistics for $\Omega_1$ and associated 68.3\% Bayesian credible intervals that exclude zero for both stars. These results depend only very weakly on the errors on the stellar parameters, which reinforces the robustness of our conclusions.

These results agree very well with other estimates of latitudinal differential rotation obtained either by spectropolarimetry, spectroscopy, or photometry. In particular, they seem to follow the $\Delta\Omega - \teff$ relationship suggested by \citet{Barnes05a}. They are of particular significance in that they represent the first conclusive detection for solar analogues, however. So far, mostly young active stars, often still in the post-main sequence stage, have yielded convincing measurements. Together with the results of \citet{Benomar18}, this work has opened the door for the study of differential rotation in main-sequence stars using asteroseismology. More precisely, it demonstrates the feasibility of such a detection for solar analogues. Studying such objects is important since their physical states are similar to those of the Sun. Therefore, theoretical models developed for this latter are likely to still apply to these stars. A fascinating perspective would thus be to gather similar data for other solar analogues and/or solar twins, using instruments such as SONG, TESS, and PLATO, in order to be able to constrain existing theoretical models for differential rotation and even dynamo.

\begin{acknowledgements}

We thank the referee for their careful reading of the manuscript. The authors thank C.~Hanson for interesting discussions. This material is based upon work supported by the NYU Abu Dhabi Institute under grant G1502. Funding for the Stellar Astrophysics Centre is provided by The Danish National Research Foundation (Grant DNRF106). The research was supported by the ASTERISK project (ASTERoseismic Investigations with SONG and Kepler) funded by the European Research Council (Grant agreement no.: 267864). In memory of Michael~J. Thompson.

\end{acknowledgements}

\bibliography{ref}

\appendix

\section{Modelling 16~Cyg~A and B}\label{app:model}

\begin{table}[h!]
\center
\caption{Non-seismic observational properties for the two stars of the 16~Cyg system.}
\label{tab:obs}      
\centering                                      
\begin{tabular}{lcccc}
\hline\hline
\\[-3.mm]
Star &$\teff$ (K)& [Fe/H] & $L/L_{\odot}$ & $R/R_{\odot}$ \\
\hline
\\[-3.mm]
A & $5825\pm 50$  & $0.10\pm0.09$  & $1.56\pm0.05$ & $1.22\pm0.02$ \\
B & $5750\pm 50$  & $0.05\pm0.06$  & $1.27\pm0.04$ & $1.12\pm0.05$ \\
\hline
\end{tabular} 
\end{table}

A common exercise, albeit challenging, in stellar physics consists of estimating stellar parameters such as the mass, $M_{\star}$, the age, $\stage$, the initial composition given by the initial hydrogen-mass fraction, $X_0$, and initial metallicity, $Z_0$, and the mixing-length parameter, $\alpha$. In the following, we group them in a single vector $\thetastv = (M_{\star},\stage,X_0,Z_0,\alpha)$. The parameters are real numbers. We not only wish to estimate the value of $\thetastv$, but also the uncertainty in this value due to the errors on the observational constraints.

\begin{table}[h!]
\begin{center}
\caption{Lower and upper bounds used for the prior uniform densities for each stellar parameter.}
\label{tab:priors}
\begin{tabular}{@{}lcc@{}}
\hline
\hline
Parameter&  Lower bound & Upper bound\\
\hline
$M$ (M$_{\odot}$)      & 0.7& 1.25\\
$\stage$ (Gyr)        &  1& 13\\
$Z_0$                 & 0.010&  0.027\\
$X_0$                 & 0.525& 0.750\\
$\alpha$              & 1.& 3.\\
\hline
\end{tabular}
\end{center}
\end{table}
The observational data can come from many different sources, spectroscopy, photometry, interferometry, or astrometry. In the case of 16~Cyg~A and B, we used an effective temperature and surface metallicity, $T_{\mathrm{eff}}$ and [Fe/H], derived from high-precision spectroscopy measurements \citep{Ramirez09}. The radius was obtained using interferometry \citep{White13}. The luminosity was derived from the astrometric Hipparcos parallax \citep{vanLeeuwen07}. The seismic data were processed from the same \emph{Kepler} time series as we used in this study, published by \cite{Davies15}. We did not use the individual frequencies directly to constrain our model because this demands the introduction of heuristic surface correction to our theoretical model. Rather, we used the frequency ratios 
  \begin{align}
    \displaystyle
    r_{01}(n) &= \frac{\nu_{n-1,0} - 4\nu_{n-1,1} + 6\nu_{n,0} - 4\nu_{n,1} + \nu_{n+1,0}}{8(\nu_{n,1} - \nu_{n-1,1})},\\
    r_{02}(n) &= \frac{\nu_{n,0} - \nu_{n-1,2}}{\nu_{n,1} - \nu_{n-1,1}} ,
\label{eq:seprat}
\end{align}
defined by \citet{Roxburgh03}. These are expected to be far less sensitive to the surface and thus stand out as adequate quantities for model fitting \citep{Bazot13,SA13}. The non-seismic observational constraints are listed in Table~\ref{tab:obs}. In the following, the observations are grouped in a vector $\Xstv = (T_{\mathrm{eff}},\mathrm([Fe/H],R,L,\boldsymbol{r_{01}},\boldsymbol{r_{02}}))$, with $\boldsymbol{r_{01}} = (r_{01}(n_{01,1}),\dots,r_{01}(n_{01,N}))$ and $\boldsymbol{r_{02}} = (r_{02}(n_{02,1}),\dots,r_{02}(n_{02,M}))$ (the indices $n_{01,i}$ and $n_{02,i}$ represent the mode orders for which the corresponding ratio can be evaluated from the observed oscillation frequencies).
\begin{table*}
\center
\caption{Stellar parameters (and helium-mass fraction) inferred using the ASTEC stellar evolution code for 16~Cyg~A and B. For each star, the first line gives the global MAP estimate. The second line is the MAP estimate for each marginalised density. The third line gives the posterior mean estimate.}

\label{tab:params}      
\centering                                      
\begin{tabular}{lcccccc}          
\hline\hline
\\[-3.mm]
Star &$M/M_{\odot}$ & $\stage$ (Gyr) & $X_0$ & $Z_0$ & $\alpha$ & $Y_0$\\
\hline
\\[-3.mm]
& $1.06$ & $6.74$ & $0.694$ & $0.0240$ & $2.12$ & $0.282$\\[1mm]
\multirow{1}{*}{16~Cyg~A}& $1.07^{+0.02}_{-0.02}$ & $6.70^{+0.23}_{-0.17}$ & $0.698^{+0.013}_{-0.013}$ & $0.0237^{+0.0019}_{-0.0018}$ & $2.12^{+0.09}_{-0.07}$ & ${0.279}^{+0.012}_{-0.013}$\\[1mm]
& 1.07 (0.02) & 6.73 (0.19) & 0.698 (0.013) & 0.0238 (0.0018) & 2.13 (0.08) & 0.278 (0.012)\\
\hline
\\[-2.mm]
&$1.05$ & $6.73$ & $0.711$ & $0.0234$ & $2.09$ & $0.266$\\[1mm]
\multirow{1}{*}{16~Cyg~B} & $1.05^{+0.02}_{-0.02}$ & $6.63^{+0.20}_{-0.19}$ & $0.706^{+0.014}_{-0.015}$ & $0.0242^{+0.0022}_{-0.0019}$ & $2.12^{+0.08}_{-0.07}$ & ${0.270}^{+0.013}_{-0.013}$\\[1mm]
&1.05 (0.02) & 6.64 (0.18) & 0.706 (0.014) & 0.0245 (0.0020) & 2.12 (0.07) & 0.270 (0.013)\\
\hline
\end{tabular}
\\[3.mm]
\end{table*}

Many difficulties are present when fitting stellar data with theoretical models. In brief, the main challenges stem from the facts that the theoretically evaluated observables depend non-linearly on $\thetastv$ and that the computational cost of stellar models is relatively high. The former issue implies that sophisticated methods of computational statistics may be required to solve the estimation problem. The latter problem makes it difficult to use such methods.

We are interested in obtaining approximations to the joint posterior density of the stellar parameters, $\thetastv \sim \pi_{\thetastv\vert\Xstv}$. We obtained an expression for this density using Bayes' formula (\ref{eq:posterior}) for $\thetastv$ and $\Xstv$. In this context, the likelihood was obtained by assuming that the observations are the sum of a deterministic and stochastic component
\begin{equation}\label{eq:statadd}
  \Xstv = \mathcal{S}(\thetastv) + \epsilonv,
\end{equation}
with $\mathcal{S}(\thetastv)$ a mapping from the space of parameters to the space of observations that represents the stellar evolution code, and $\epsilonv$ the realisation of a random vector.

We assumed that the uncertainties on $T_{\mathrm{eff}}$, [Fe/H], $L$ and $R$ are Gaussian with respective standard deviations $\sigma_{T_{\mathrm{eff}}}$, $\sigma_{\mathrm{[Fe/H]}}$, $\sigma_{L}$ , and $\sigma_{R}$ the observational uncertainties. The components of $\boldsymbol{r_{01}}$ and $\boldsymbol{r_{02}}$ are correlated, therefore we treated these vector as two separate multivariate Gaussian densities $\mathcal{N}(\boldsymbol{\mu_{01}},\boldsymbol{\Sigma_{01}})$ and $\mathcal{N}(\boldsymbol{\mu_{02}},\boldsymbol{\Sigma_{02}})$. The covariance matrices were estimated numerically. Independent samples were generated for each frequency used for the evaluation of the components of $\boldsymbol{r_{01}}$ and $\boldsymbol{r_{02}}$, and were used to obtain samples for both random vectors. Using these samples, evaluating $\boldsymbol{\Sigma}_{01}$ and $\boldsymbol{\Sigma}_{02}$ is straightforward. The resulting likelihood is therefore
\begin{widetext}
  \begin{equation}\label{eq:likeli-stellar}
  \begin{split}
  \displaystyle
  \pi(\Xstv\vert\thetastv) \propto \exp \Bigg[ -\frac{1}{2} \Bigg( \frac{(T - \langle T_{\mathrm{eff}} \rangle)^2}{\sigma^2_{T_{\mathrm{eff}}}} + \frac{(\mathrm{[Fe/H]} - \langle \mathrm{[Fe/H]} \rangle)^2}{\sigma^2_{\mathrm{[Fe/H]}}} + & \frac{(L - \langle L \rangle)^2}{\sigma^2_L} + \frac{(R - \langle R \rangle)^2}{\sigma^2_R}  \\
      & + (\boldsymbol{r_{01}} - \boldsymbol{\mu_{01}})^T\boldsymbol{\Sigma_{01}}(\boldsymbol{r_{01}} - \boldsymbol{\mu_{01}}) + (\boldsymbol{r_{02}} - \boldsymbol{\mu_{02}})^T\boldsymbol{\Sigma_{02}}(\boldsymbol{r_{02}} - \boldsymbol{\mu_{02}}) \Bigg)\Bigg].
  \end{split}
    \end{equation}
\end{widetext}
Here the average quantities, denoted by $\langle . \rangle$, are the observed quantities. We note the difference of functional form between Eq.~(\ref{eq:likelihood}) and Eq.~(\ref{eq:likeli-stellar}). This stems from the difference in the underlying statistical model.

The first term in Eq.~(\ref{eq:statadd}) does not have an analytic closed form. In order to express it, we must solve the equation for stellar structure and pulsations numerically. This was achieved using the Aarhus Stellar Evolution Code (ASTEC) for the former and {\tt adipls} for the latter. We assumed spherical symmetry and no magnetic field.
The opacities were obtained from OPAL tables \citep{Iglesias96}, with low-$T$ opacities from \citet{Ferguson05}, and the equation of state was interpolated from OPAL tables \citep{OPAL02}. Nuclear reaction rates were taken from the NACRE collaboration \citep{Angulo99} and supplemented by the values given in \citet{Imbriani05} for the $^{14}$N(p,$\gamma$)$^{15}$O reaction. Convection was treated using the prescription from \citet{BV58} for the mixing-length theory, the mean-free path of the fluid elements being proportional to the pressure scale-height.

\begin{figure}[h!]
   \centering
   \includegraphics[width=.45\textwidth]{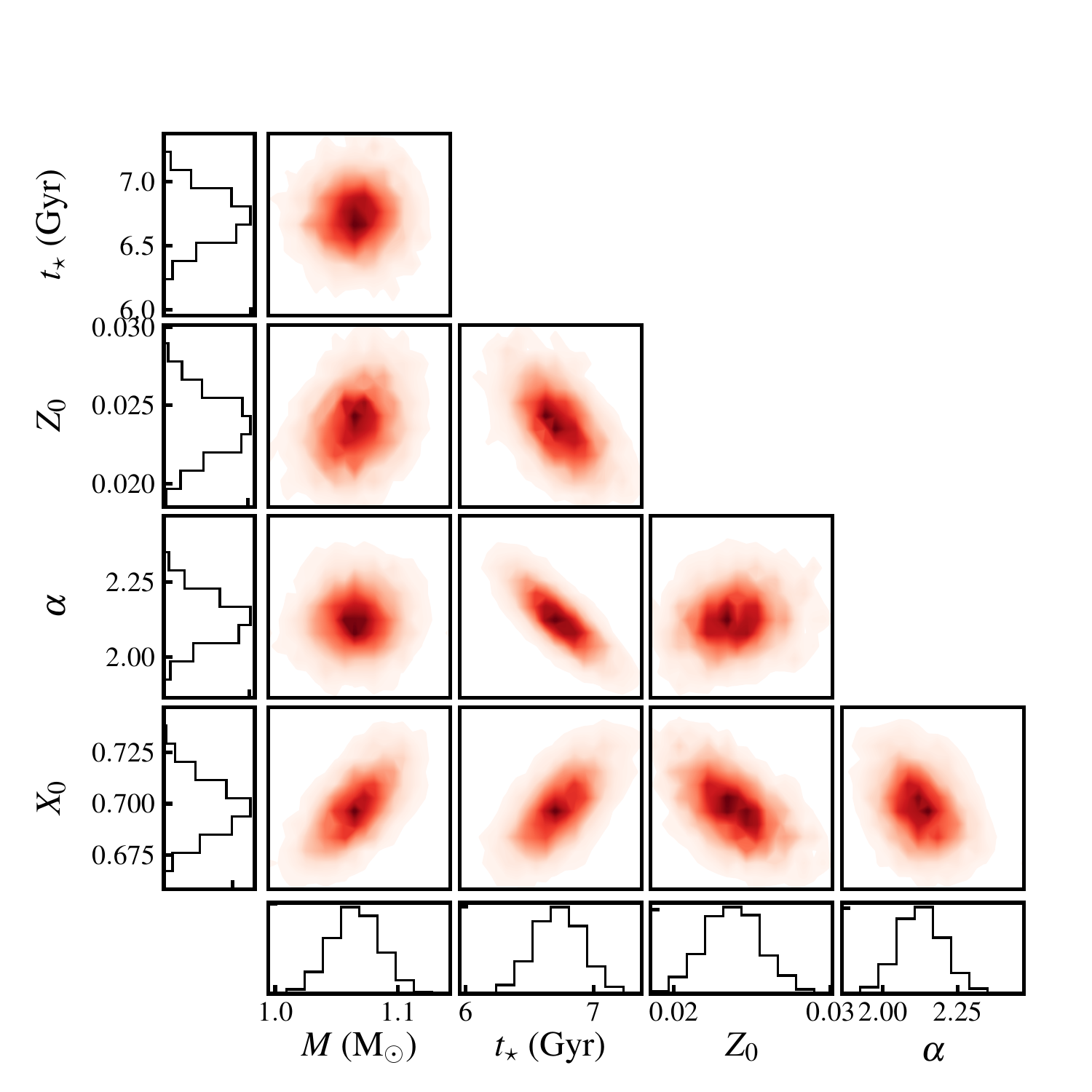}
   \caption{Marginal densities for the stellar parameters $M$, $\stage$, $X_0$, $Z_0$ , and $\alpha$ of 16~Cyg~A. The central panels show the joint marginal densities of the paired parameters. Individual marginal densities are plotted in the side panels. 
   }
         \label{fig:joint-16CygA-stellar}
   \end{figure}
\begin{figure}[h!]
   \centering
   \includegraphics[width=.45\textwidth]{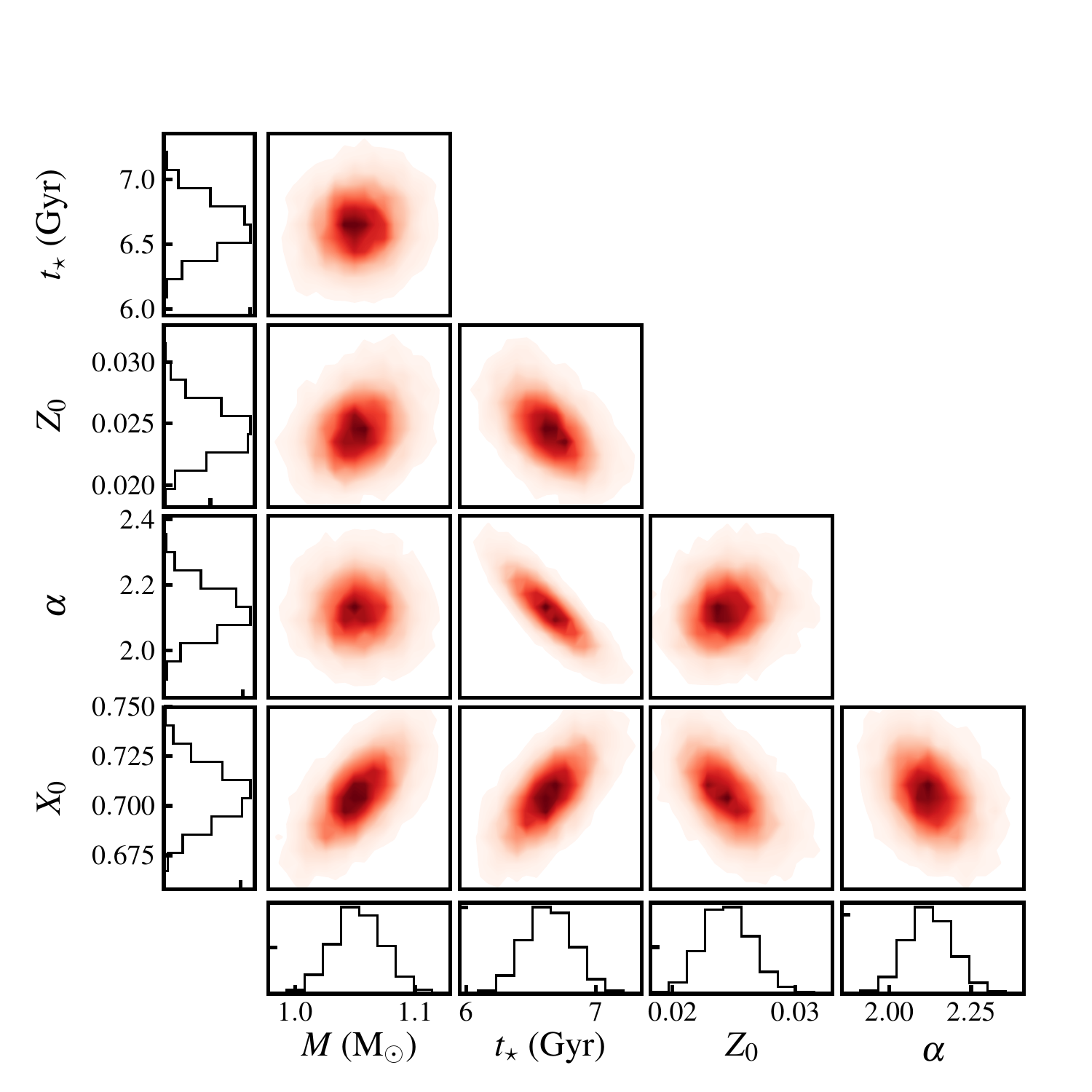}
   \caption{Marginal densities for the stellar parameters $M$, $\stage$, $X_0$, $Z_0$ , and $\alpha$ of 16~Cyg~B. The central panels show the joint marginal densities of the paired parameters. Individual marginal densities are plotted in the side panels.. 
   }
         \label{fig:joint-16CygB-stellar}
   \end{figure}

The stellar parameters were considered to be independent. Therefore, the prior can be written $\pi(\thetastv) = \pi(M_{\star})\pi(\stage)\pi(X_0)\pi(Z_0)\pi(\alpha)$. Priors were chosen as uniform because we do not have previous measurements on any of them. All these quantities are positive, therefore the lower bounds of these prior densities should be non-negative. To set both upper and lower bounds, we used the estimates obtained by \citet{Metcalfe15} as first initial guesses. These were obtained using almost the exact same data. There are good indications of the range in which we expect the significant probability mass to be found. We then refined the boundaries on our priors using successive trial-and-error stages. This was done in order to avoid to sharp cuts in the domain of definition of the posterior density. This could indeed lead to numerical issues when sampling from a posterior density using an MCMC algorithm. The only notable exception to this procedure concerns the initial hydrogen-mass fraction, which cannot be higher than 0.75, which is its value after the primordial nucleosynthesis \citep[see][for a discussion]{Bazot12,Bazot16}. The priors used in our statistical model are given in Table~\ref{tab:priors}.

   The posterior density was sampled using an MCMC algorithm. The details of the algorithm can be found in \citet{Bazot18}. It was run on ten independent chains. Each chain was {}heated{} with {}a temperature{} $T > 1$, so that we initially sampled a posterior density of the form $\pi(\thetastv)\pi^{1/T}(\Xstv\vert\thetastv)$. This procedure, known as simulated annealing according to \citet{Kirkpatrick83}, allowed us to sample the space of parameters for densities with much weaker variations than the original target. Therefore, proposals of an MCMC algorithm will tend to be accepted more often after heating. This sometimes helps avoiding that a Markov chain becomes stuck far from the real solution that is sought for, in low-probability regions. It is therefore possible to run a preliminary sequence of MCMC runs with decreasing temperatures (with the last chain having $T=1$), in order to identify the regions of high probability. The parameter $T$ was assigned a decreasing sequence $2^{n_T}$ with $n_T$ an integer such that $0 \leq n_T \leq 6$. The number of iteration was 1000 for $n_T = 6$ and 200 for $1 \leq n_T \leq 5$. For $n_T = 0,$ the chains were run until acceptable convergence was obtained. The chains were initialised at $n_T = 6$ using an overdispersed density, crudely estimated from a short test run. At subsequent stages, all chains were initialised at the MAP value of the previous one. Convergence was tested using several diagnostics: the cumulative mean, variance, and the $r$ ratio defined by \citet{Gelman92}.

The two-dimensional and one-dimensional posterior densities for the stellar parameters of 16~Cyg~A and B are shown in Figs.~\ref{fig:joint-16CygA-stellar} and ~\ref{fig:joint-16CygB-stellar}. Corresponding estimates are also given in Table~\ref{tab:params}. We also give the initial helium-mass fraction, $Y_0$, which is an often-used parameter in the literature. The MAP values computed from the five-dimensional joint density are given in the first line. They are given without uncertainties. We also give the MAP values obtained from each marginalised one-dimensional posterior, as shown in the side panels of Figs.~\ref{fig:joint-16CygA-stellar} and \ref{fig:joint-16CygB-stellar}. In this case, we produced an associated credible interval. The latter is defined as the smallest interval of probability mass 0.683 that encompasses the MAP. Finally, we also give the posterior mean and the posterior variance. All three estimates agree with each other. The estimates so obtained are in fair agreement with those of \citet{Metcalfe15}, even though a detailed comparison with this work is well beyond the scope of our study. The uncertainty on the parameters is remarkably low. In particular, the age is known with a precision of roughly 200~Myr. 

\end{document}